\documentclass[apj,a4]{emulateapj}

\shorttitle{}
\shortauthors{}

\begin{document}

\title{No Evidence for Evolution in the Far-Infrared-Radio Correlation out to $z$ $\sim$ 2 in the ECDFS}

\author{Minnie Y. Mao}
\affil{Infrared Processing and Analysis Center, California Institute of Technology, Pasadena CA 91125, USA. minnie.mao@csiro.au}
\affil{School of Mathematics and Physics, University of Tasmania, Private Bag 37, Hobart, 7001, Australia}
\affil{CSIRO Astronomy and Space Science, PO Box 76, Epping, NSW, 1710, Australia}
\affil{Australian Astronomical Observatory, PO Box 296, Epping, NSW, 1710, Australia}

\author{Minh T. Huynh}
\affil{Infrared Processing and Analysis Center, California Institute of Technology, Pasadena CA 91125, USA}
\affil{International Centre for Radio Astronomy Research, M468, University of Western Australia, Crawley, WA 6009, Australia}
\author{Ray P. Norris}
\affil{CSIRO Astronomy and Space Science, PO Box 76, Epping, NSW, 1710, Australia}

\author{Mark Dickinson}
\affil{National Optical Astronomy Observatory, 950 North Cherry Avenue, Tucson, AZ 85719, USA}

\author{Dave Frayer}
\affil{National Radio Astronomy Observatory, PO Box 2, Green Bank, WV 24944, USA}

\author{George Helou}
\affil{Infrared Processing and Analysis Center, California Institute of Technology, Pasadena CA 91125, USA}

\and

\author{Jacqueline A. Monkiewicz}
\affil{School of Earth and Space Exploration, Arizona State University, Tempe, AZ 85287, USA}

\begin{abstract}

We investigate the 70$\,\mu$m Far-Infrared Radio Correlation (FRC) of star-forming galaxies in the Extended Chandra Deep Field South (ECDFS) out to $z$ $>$ 2. We use 70$\,\mu$m data from the Far-Infrared Deep Extragalactic Legacy Survey (FIDEL), which comprises the most sensitive ($\sim$0.8\,mJy rms) and extensive far-infrared deep field observations using MIPS on the Spitzer Space Telescope, and 1.4\,GHz radio data ($\sim$8$\,\mu$Jy\,beam$^{-1}$ rms) from the VLA. In order to quantify the evolution of the FRC we use both survival analysis and stacking techniques which we find give similar results. We also calculate the FRC using total infrared luminosity and rest-frame radio luminosity, $q_{TIR}$, and find that $q_{TIR}$ is constant (within 0.22) over the redshift range 0 - 2. We see no evidence for evolution in the FRC at 70$\,\mu$m, which is surprising given the many factors that are expected to change this ratio at high redshifts.

\end{abstract}


\section{Introduction}

The correlation between the far-infrared (FIR) and radio emission for star-forming galaxies in the local Universe was first observed by \citet{vanderKruit71,vanderKruit73} and is the tightest and most universal correlation known among global parameters of galaxies \citep{Helou93}. The correlation is linear, spans five orders of magnitude of bolometric luminosity and has been shown to hold for a wide range of Hubble types \citep{deJong85, Helou85, Condon92, Yun01}. 

The FIR-radio correlation (FRC) has been attributed to the presence of young, high-mass (M $>$ 8M$_\odot$) stars. The FIR emission arises from the absorption by dust of the UV emission and subsequent reradiation of the energy at IR wavelengths. The radio emission is dominated by non-thermal synchrotron emission from cosmic ray electrons which are accelerated by supernovae shocks. There is also a thermal component which arises from free-free emission from ionized hydrogen in H{\scriptsize II} regions, but this contributes only $\sim$10$\%$ of the radio emission at lower frequencies \citep[$<$5\,GHz,][]{Condon92}, and becomes more significant at higher frequencies. 

The ``calorimeter theory'' \citep{Volk89} suggests that the FRC holds because galaxies are both electron calorimeters and UV calorimeters so the total radio and IR outputs remain proportional independent of variations within the galaxy. This theory however, requires that galaxies are optically thick to UV light from the young high-mass stars, and thus does not hold for optically-thin galaxies. \citet{Helou93} proposed the non-calorimetric ``optically thin'' scenario involving a correlation between disk scale height and the escape scale length for cosmic ray electrons, while \cite{Bell03} concludes that the linearity of the FRC is a conspiracy as the star-formation rate is underestimated for low-luminosity galaxies at both radio and infrared frequencies. However, these models typically leave out proton losses and non-synchrotron cooling \citep{Lacki10}. Ultimately, the physical origin of the FRC is still not clear.

The far-reaching nature of the FRC has made it a valuable diagnostic. Some examples of its application include: using the FRC to identify radio-loud AGN \citep{Donley05, Norris06}; using the FRC to define the radio luminosity/SFR relation \citep{Bell03}; and, at higher redshifts, using the FRC to estimate distances to submillimeter galaxies without optical counterparts \citep[e.g.,][]{Carilli99}. Consequently it is of great importance to determine whether the FRC holds at high redshifts. 

The FRC may fail at high redshifts for a number of reasons. Electrons are expected to lose energy by inverse Compton interactions with the cosmic microwave background, whose energy density scales as (1+z)$^4$, implying a lower level of radio emission at higher redshifts. Moreover, synchrotron emission is proportional to the magnetic field strength squared, so evolution of magnetic field strength should affect the FRC at higher redshifts \citep[e.g.,][]{Murphy09}. Changes in the spectral energy distributions (SEDs) may also be expected due to evolution in dust properties and metallicity \citep[e.g.,][]{Amblard10, Hwang10, Chapman10}. Nonetheless, current studies show no firm evidence for evolution in the FRC \citep[e.g.,][]{Garrett02, Appleton04, Seymour09, Bourne10, Ivison10a, Ivison10b, Sargent10a, Sargent10b, Huynh10}. 

\emph{Herschel} was launched in May 2009 and can probe the FIR to submillimeter regime from 55$\,\mu$m to 671$\,\mu$m, deeper than ever before \citep{Pilbratt10}. Most recently, \citet{Jarvis10} and \citet{Ivison10b} used data from Herschel and found no evidence for evolution in the FRC out to $z$ = 0.5 and $z$ = 2 respectively.

This paper studies the dependence of the FRC on redshift using deep 70$\,\mu$m data from the Spitzer Space Telescope and 1.4\,GHz data from the Very Large Array (VLA). This work differs from previous studies as we are using FIDEL (Far-Infrared Deep Extragalactic Legacy Survey) 70$\,\mu$m data, which is the deepest 70$\,\mu$m data taken to date. FIDEL reaches a point source rms sensitivity of 0.8\,mJy at 70$\,\mu$m, making it far more sensitive than previous studies of the FIR at 70$\,\mu$m such as \citet{Sargent10b} whose data reached a point source rms sensitivity of 1.7\,mJy.

We focus on the 70$\,\mu$m data as this band probes closer to the dust emission peak ($\sim$100$\,\mu$m) than, for example, 24$\,\mu$m. Furthermore, for $z$ $<$ 3, the 70$\,\mu$m band is not affected by emission from polycyclic aromatic hydrocarbons (PAHs; 7 - 12 $\mu$m). While \citet{Bourne10} also study the FRC using FIDEL data, their work was entirely based on stacking analysis. This work is the first to use such deep 70$\,\mu$m data to study the evolution of the FRC based on individual sources.

The data are described in Section 2, Section 3 describes the data analysis while Section 4 presents our results and analysis. This paper uses H$_0$ = 71 km s$^{-1}$ Mpc$^{-1}$, $\Omega$$_M$ = 0.27 and $\Omega$$_\Lambda$ = 0.73.

\section{Data}
\subsection{FIDEL}
FIDEL, the Far-Infrared Deep Extragalactic Legacy Survey, is a legacy science program (PI: Dickinson) which comprises the most sensitive and extensive FIR deep field observations using the Multiband Imaging Photometer for SIRTF (MIPS) on the Spitzer Space Telescope. Characteristics of the FIDEL data are described in detail in \citet{Magnelli09} and other papers. FIDEL observed three fields: ECDFS, EGS and GOODS-North. Observations were taken at 3 bands: 24$\,\mu$m, 70$\,\mu$m and 160$\,\mu$m, focusing specifically on the 70$\,\mu$m data. This paper concentrates only on the 30$'$ $\times$ 30$'$ ECDFS field centered on 03 32 00, -27 48 00 (J2000). The observations were designed to achieve roughly uniform sensitivity at 70$\,\mu$m across most of the ECDFS, although the FIDEL data include a somewhat deeper central region with data from a GO program (PI: Frayer).  The mean 70$\,\mu$m exposure time over the ECDFS is approximately 6600s, yielding an RMS point source sensitivity of approximately 0.8\,mJy. The 24$\,\mu$m exposure time varies considerably more over the field, from 5000 to 35000 seconds over most of the ECDFS, with an average exposure time of approximately 16000 seconds.  The RMS point source sensitivity at 24$\,\mu$m thus also varies, but is typically in the range 8 to 14 $\mu$Jy over most of the field.  FIDEL data were processed using the Mosaicking and Point-source Extraction \citep[MOPEX,][]{Makovoz05} package to form the mosaicked images. The final 70$\,\mu$m mosaic has a pixel scale of 4.0$\,''$/pixel and a point response function (PRF) with an 18$''$ \emph{FWHM}. The final 24$\,\mu$m mosaic has a pixel scale of 1.2$\,''$/pixel and a PRF with a 5.9$''$ \emph{FWHM}.

\subsection{Ancillary data}
\subsubsection{Radio data}
The ECDFS has been observed at 1.4\,GHz by both the VLA and the ATCA
\citep{Norris06, Kellermann08, Miller08}.  Here we use the
\citet{Miller08} radio data due to its high angular resolution and
sensitivity over our field. The radio data encompass a 34$'$
$\times$ 34$'$ region, centered on 03 32 28.0, -27 48 30.0 (J2000). The data
have a typical rms sensitivity of 8$\,\mu$Jy per 2.8$''$ $\times$
1.6$''$ beam. The catalog contains 464 sources above a 7 sigma cutoff.

\subsubsection{Redshift data}
We obtained both spectroscopic and photometric redshift data from COMBO-17 \citep{Wolf04}, MUSYC \citep{Gawiser06, Cardamone10}, GOODS \citep{Balestra10} and ATLAS \citep{Mao09}. 

COMBO-17 has photometric data in 17 passbands from 350$\,$nm to 930$\,$nm for 63501 objects in the ECDFS. The photometric redshifts are most reliable for sources with R $\le$ 24 \citep{Wolf04}.

MUSYC (Multiwavelength Survey by Yale-Chile) has photometric data in the ECDFS. \citet{Cardamone10} combined photometric data from the literature with new deep 18-medium-band photometry into a public catalog of $\sim$80000 galaxies in ECDFS, from which they computed the photometric redshifts. The photometric redshifts are most reliable for sources with R $\le$ 25.5. \citet{Cardamone10} also compile a spectroscopic redshift catalog of 2551 galaxies from the the literature \citep[e.g.,][]{Balestra10, Vanzella08, LeFevre04}. 

\citet{Balestra10} observed the GOODS-South field (within the ECDFS) using VIMOS to obtain spectroscopic redshifts. Their campaign used two different grisms to cover different redshift ranges and used 20 VIMOS masks. They combined their resulting redshifts with those available in the literature to produce a catalog containing 7332 spectroscopic redshifts. Quality flags were provided for all the redshifts and we took only those redshifts that had quality flags of ``secure'' and ``likely'' yielding a catalog of 5528 spectroscopic redshifts. Although \citet{Cardamone10} includes \citet{Balestra10} data in their compilation of spectroscopic redshifts, they only include a subset of the data.

\citet{Mao09} are undertaking a program of redshift determination and source classification of all ATLAS (Australia Telescope Large Area Survey) radio sources with AAOmega \citep{Sharp06} on the Anglo-Australian Telescope (AAT). Using redshifts from both the literature and their own campaign they have a total of 261 spectroscopic redshifts in ECDFS and its surrounding region. 

\subsubsection{X-ray data}
We use the 2~Ms Chandra Deep Field South X-ray catalogs from \citet{Luo08} to identify AGN in our 70$\,\mu$m catalog (Section \ref{agndiag}). This is one of the most sensitive X-ray surveys ever performed and detects 462 sources in 436 arcmin$^2$. While the X-ray data cover a smaller region than the FIDEL data, we use these data to eliminate some AGN from our sample (Section \ref{agndiag}). 

\section{Data Analysis}

\subsection{FIDEL}
\subsubsection{70$\,\mu$m}

The 70$\,\mu$m catalog was produced with the Astronomical Point Source Extraction (APEX) module within the MOPEX package. APEX subtracts the local background by calculating the median in a region, which we set to 34 $\times$ 34 pixels, surrounding each pixel, and removing the 100 brightest pixels. Peak values with a S/N greater than three were fitted using the PRF. The 3-sigma catalog extracted using APEX contained 515 sources. 

\subsubsection{24$\,\mu$m}
A similar source detection and extraction process to the 70$\,\mu$m data were used for the 24$\,\mu$m data. However a four sigma cutoff was used to reduce the number of spurious sources. Visual inspection was required to remove spurious sources due to artifacts surrounding bright objects. The final catalog of 24$\,\mu$m sources in the $\sim$30$'$ $\times$ 30$'$ region of interest contained 5319 sources.

\subsubsection{Final catalog}

\begin{figure*}
\begin{center}
\includegraphics[angle=0, scale=0.8]{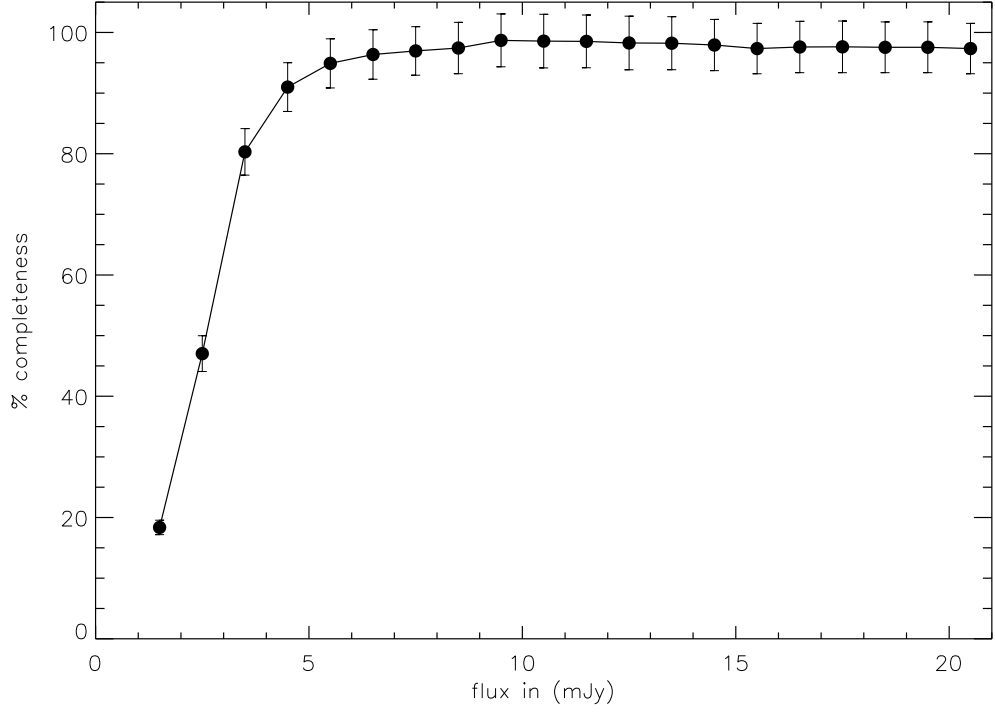}
\end{center}
\caption{The percentage completeness plotted against flux density. Our catalog is almost 100$\%$ complete at 9\,mJy, and 50$\%$ complete at $\sim$2.5\,mJy. }\label{completeness}
\end{figure*}

To test the completeness of the 70$\,\mu$m catalog, we performed Monte-Carlo simulations. A simulated source was injected at a random location in the 70$\,\mu$m image, and source extraction was performed using the same technique as for the production of the catalog. The input flux density of the simulated source varied over the range of fluxes in the ``real'' image. This process was repeated 10000 times and we tracked the simulated source recovery rate, which enables us to estimate the overall completeness over the entire image. Figure \ref{completeness} presents the completeness as a function of flux density. Our catalog is almost 100$\%$ complete at 9\,mJy, and 50$\%$ complete at $\sim$2.5\,mJy. This high level of completeness at 2.5\,mJy ($\sim$3 sigma) is probably due to the 70$\,\mu$m image not being completely uniform. The 10 $\times$ 10 arcminute region in the center has an RMS point source sensitivity of $\sim$0.6\,mJy, so 2.5\,mJy is a $>$4 sigma limit for $\sim$11\% of the image area. The completeness plot does not reach 100$\%$ due to the random positioning of the simulated source. If, by chance, the simulated source is injected upon a real source, the simulated source is not recovered as an individual source, and instead, the ``real'' source is recovered with a larger flux density.

\begin{figure*}
\begin{center}
\includegraphics[angle=0, scale=0.8]{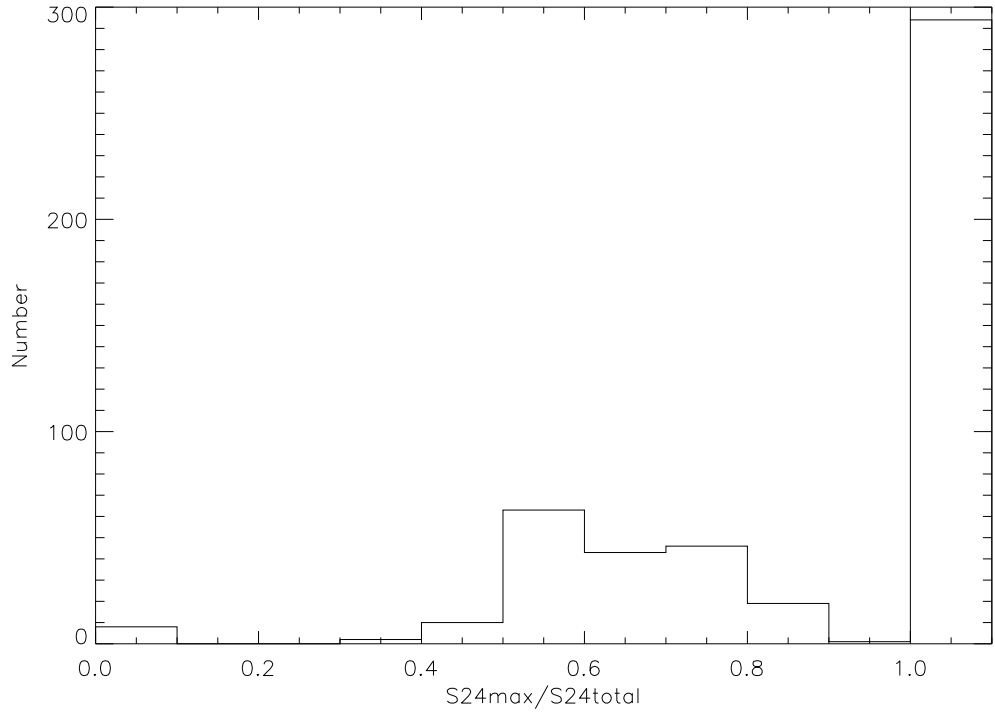}
\end{center}
\caption{Histogram of the ratio of the maximum 24$\,\mu$m flux density over the total 24$\,\mu$m flux density within a 9$''$ radius. Sources with a ratio less than 0.8 were deemed deblend candidates.}\label{deblend}
\end{figure*}

The 70$\,\mu$m catalog of 515 sources obtained using APEX may contain confused sources, as well as spurious sources. We determined which sources required deblending by comparing it with the 24$\,\mu$m catalog which has better resolution. First, we matched the 70$\,\mu$m catalog to the 24$\,\mu$m catalog using a matching radius of 9$''$ (the half-width at half maximum of the PRF at 70$\,\mu$m). We then calculated the ratio of the flux density of the brightest 24$\,\mu$m source to the total flux density of all 24$\,\mu$m sources within the 9$''$ matching radius (see Figure \ref{deblend}). Those sources for which the  ratio is one would not require deblending. Using arguments similar to those of \citet{Pope06}, we assume that if a source is bright at 24$\,\mu$m, it will also be bright at 70$\,\mu$m. Consequently, if the ratio was greater than 0.8 (that is, if the brightest 24$\,\mu$m source contributed $>$80$\%$ of the total flux density in the matching radius), the 70$\,\mu$m emission was determined to be from the brightest 24$\,\mu$m source. If the ratio is less than 0.8, we considered the 70$\,\mu$m source as a candidate for deblending. Using these criteria we determined that 164 sources were deblend candidates. In principle, it is possible that for sources with a ratio of greater than 0.8, we are overestimating the 70$\,\mu$m flux density by up to 20\%. An overestimation of the 70$\,\mu$m flux density may result in a higher FRC, but this effect is much less than $\Delta q_{70}$ = 0.1.


The sources were deblended using double, triple or even quadruple Gaussian fits, constraining the centers of the Gaussians to the 24$\,\mu$m positions. We visually inspected all the deblend candidates and a small number of these clearly did not require deblending as the secondary 24$\,\mu$m source was right on the edge of the 9 arcsecond radius with which we computed the deblend criteria, and were subsequently discarded as deblend candidates. In total, 143 sources were deblended. resulting in a total catalog of 658 putative sources. In the course of the visual inspection, we identified 41 spurious sources. 30 of these were sources that were faint ($<$4 $\sigma$) in the 70$\,\mu$m catalog, and did not have 24$\,\mu$m counterparts, and 11 were clearly part of the Airy ring of an adjacent, bright source. This resulted in a final catalog containing 617 70$\,\mu$m sources. 

Assuming Gaussian statistics, noise spikes  are expected to produce approximately 34 spurious sources above our 3-sigma cutoff in the 70$\,\mu$m data, and approximately 7 spurious sources above our 4-sigma cutoff in the 24$\,\mu$m data. The probability of any spurious 70$\,\mu$m source lying within a 9 arcsec radius of any spurious 24$\,\mu$m source is approximately 2\%, and so we conclude that none of the 617 sources is likely to be spurious.

We note that after deblending the 70$\,\mu$m catalog we increased the number of faint sources, which changes the completeness levels. While we have improved our completeness at the fainter flux levels, this change cannot easily be quantified.

\begin{figure*}
\begin{center}
\includegraphics[angle=-90, scale=0.6]{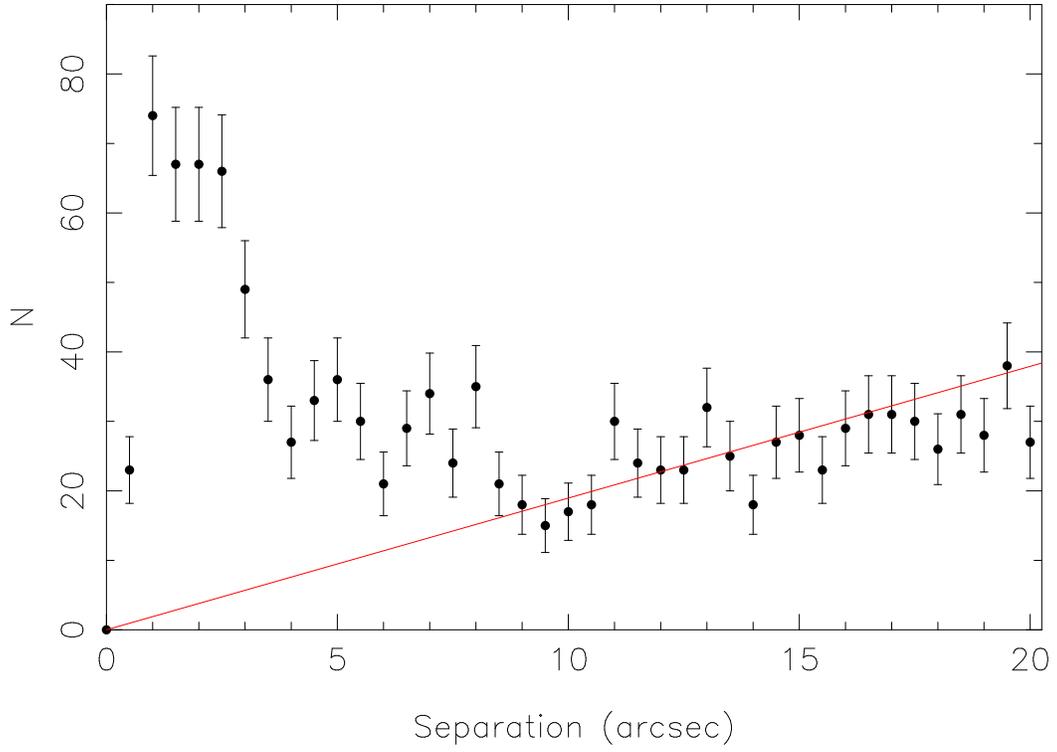}
\end{center}
\caption{Number of candidate 24$\,\mu$m counterparts of 70$\,\mu$m sources as a function of position offset. The line shows the number of matches from chance alone. }\label{match70_24}
\end{figure*}

This catalog of 617 70$\,\mu$m sources was matched to the 24$\,\mu$m catalog using a matching radius of 4$''$. Following \citet{Huynh08}, the matching radius was determined by plotting the number of candidate matches against position offset (Figure \ref{match70_24}). The matches from chance alone are determined from the source densities of the catalogs. Although Figure \ref{match70_24} suggests that a greater number of matches would be obtained using a matching radius of 8'', we choose the more conservative 4$''$ as our matching radius to minimize spurious matches.

Given the relative sensitivities of the 70$\,\mu$m and 24$\,\mu$m data it is expected that most, if not all the 70$\,\mu$m sources will have a 24$\,\mu$m counterpart. A small number (10) of strong 70$\,\mu$m sources did not have 24$\,\mu$m counterparts in the catalogs, while the 24$\,\mu$m image showed faint detections. We therefore performed aperture photometry on these faint sources and successfully extracted eight. The final catalog has only two 70$\,\mu$m sources without 24$\,\mu$m counterparts but neither of these have extreme S70/S24 ratios. 

\subsection{Radio, Optical and X-ray Counterparts}
\subsubsection{Radio Counterparts}
The 70$\,\mu$m catalog was matched to the radio catalog of \citet{Miller08}, hereafter M08, using a matching radius of 5$''$. The matching radius was determined in a similar manner to that described in the previous section. Where available, the 24$\,\mu$m position was used because the higher resolution at 24$\,\mu$m allows for better positional accuracy. This resulted in 171 radio sources matched to the 70$\,\mu$m catalog.

The M08 radio catalog has a 7 sigma cutoff. In order to increase the number of radio counterparts, and given that we have the additional information of the 70$\,\mu$m sources, we  extracted radio sources at the known 70$\,\mu$m source positions that had radio detections greater than 3 sigma. This was done by performing Gaussian fits twice, first with a fixed size (2.8$''$ x 1.6$''$ - the beam of M08 data), and then without a fixed size. The relationship between peak and integrated flux densities was determined by:
\begin{equation}
\frac{S_{int}}{S_{peak}} = \frac{\theta_{maj}\theta_{min}}{b_{maj}b_{min}},
\end{equation}
where
$S_{int}$ is the total integrated flux density,  
$S_{peak}$ is the peak flux density,
$\theta_{maj}$ is the semi-major axis of the free Gaussian,
$\theta_{min}$ is the semi-minor axis of the free Gaussian,
$b_{maj}$ is the semi-major axis of the fixed Gaussian, and
$b_{min}$ is the semi-minor axis of the fixed Gaussian.

If $S_{int}/S_{peak}$ was less than 1.2, the source was determined to be unresolved and the peak flux density for the fixed Gaussian was determined to be the final flux density. If the ratio was greater than 1.2, the source was determined to be extended, and the integrated flux was taken to be the final radio flux density. This resulted in 353 70$\,\mu$m sources with radio counterparts.

\begin{figure*}
\begin{center}
\includegraphics[angle=0, scale=0.8]{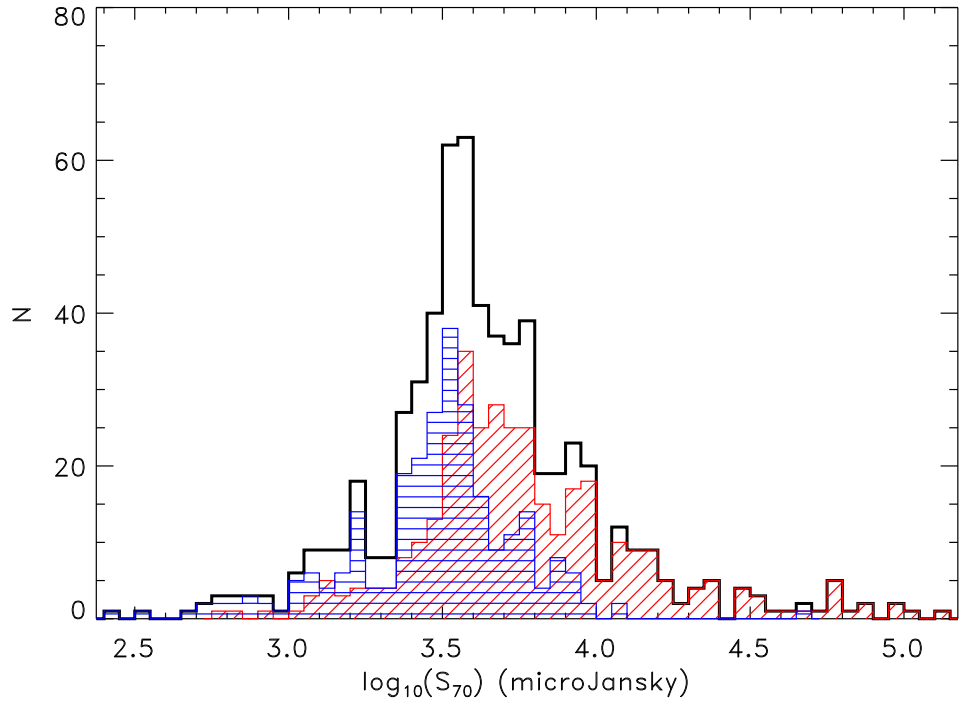}
\end{center}
\caption{Histogram of 70$\,\mu$m fluxes. The black line is the histogram for all 617 sources, while the red diagonally shaded histogram shows the 353 sources that have a radio detection and the blue horizontally shaded histogram shows the 264 sources that do not have a radio detection. }\label{s70hist}
\end{figure*}

In summary, our final 70$\,\mu$m catalog contains 617 sources, 615 of which have a 24$\,\mu$m counterpart. 353 of these have a radio detection greater than 3 sigma from \citet{Miller08}. Figure \ref{s70hist} shows the histogram of the 70$\,\mu$m flux densities. We also overplot the population of sources with and without radio counterparts. Sources with no radio counterparts have a median 70$\,\mu$m flux density of 3.3$\,$mJy while sources with radio counterparts have a median 70$\,\mu$m flux density of 5.2$\,$mJy. 

\subsubsection{Redshift Data and Infrared Luminosities}\label{zlir}
The final 70$\,\mu$m catalog was matched to the various redshift catalogs using the matching radii shown in Table \ref{redshifts}. Due to the better resolution of the radio data we used radio positions where available, followed by 24$\,\mu$m positions where available to match to the redshift catalogs. 

Where possible we matched spectroscopic redshifts to the sources. Where more than one spectroscopic redshift was available, we chose to prioritize ATLAS spectroscopic redshifts over GOODS spectroscopic data as we have access to the ATLAS spectra. The difference in redshift, $\Delta$z, was $<$0.0015 for the 12 sources in common.

Where no spectroscopic data were available we used photometric redshifts. Where both COMBO-17 and MUSYC photometric redshifts were available we chose to prioritize MUSYC data over COMBO-17 because MUSYC's photometric redshifts are derived from 32 bands as opposed to COMBO-17, which uses 17. Furthermore, MUSYC has photometric data extending into the near-infrared (JHK), which improves the accuracy of the photometric redshift.

\begin{figure*}
\begin{center}
\includegraphics[angle=0, scale=0.8]{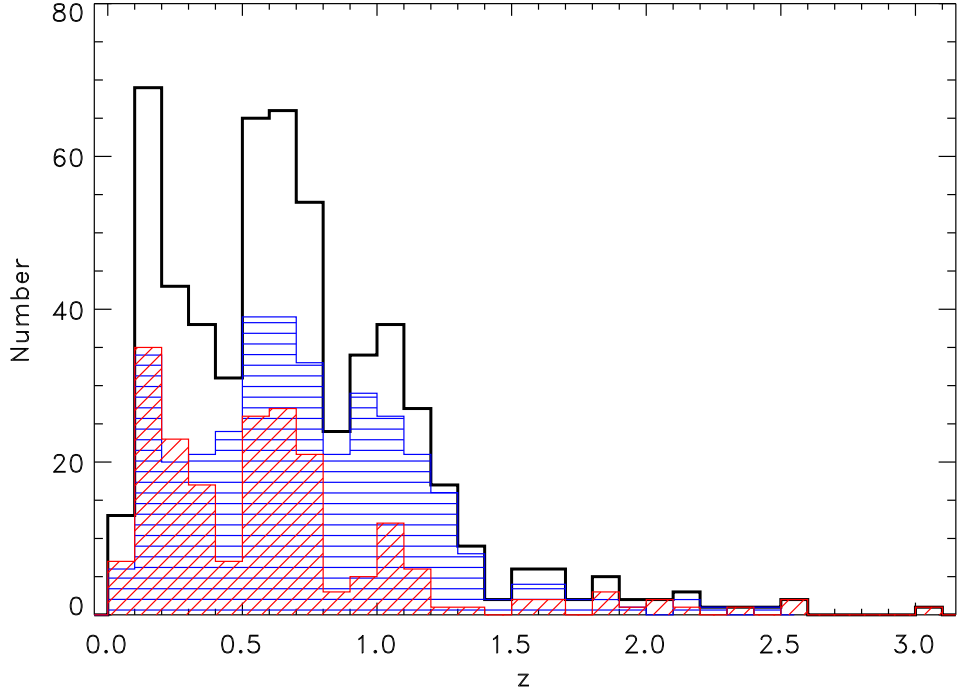}
\end{center}
\caption{Histogram of all 562 sources that have redshift information. The black line is the histogram for all sources that have redshift information while the red diagonally shaded histogram shows sources with spectroscopic redshifts and the blue horizontally shaded histogram shows sources with photometric redshifts.}\label{zhist}
\end{figure*}

In summary, 562 of the 617 (91$\%$) 70$\,\mu$m sources have redshift information, 206 (33$\%$) of which are spectroscopic (Figure \ref{zhist}). 

The total infrared luminosity (8 -- 1000$\,\mu$m) of the sources in our sample was estimated by fitting the 24$\,\mu$m and 70$\,\mu$m flux density to the SED templates of \citet{Chary01}, hereafter CE01, while letting the templates scale in luminosity. The CE01 templates show observed 24$\,\mu$m/70$\,\mu$m flux density ratios which are more representative of  $z \sim 1$ galaxies than \citet{Dale02} or \citet{Lagache03} templates. \citet{Magnelli09} found that 70\,$\mu$m data seems to provide an estimate of L$_{IR}$ that is nearly independent of the SED library used, but this is partly because 70\,$\mu$m is close to the peak emission, and therefore carries the largest fraction of FIR power. 

\begin{figure*}[bt]
\begin{center}
\includegraphics[width=14cm]{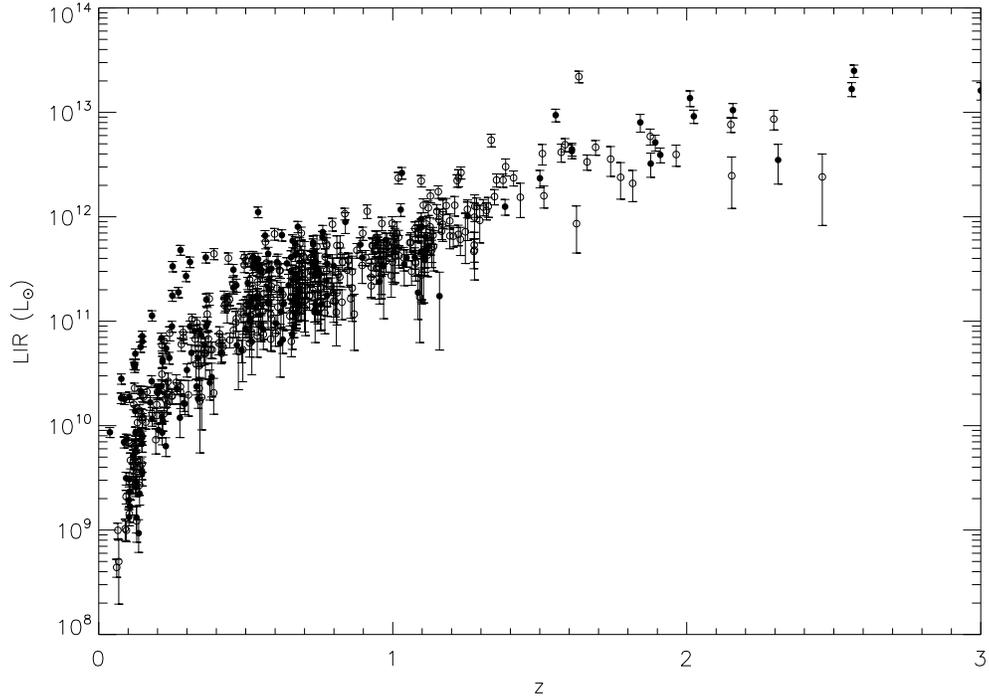}
\end{center}
\caption{The total IR luminosity (8 -- 1000 $\mu$m) as a function of redshift for all IR sources with redshift information (562/617). Sources with spectroscopic redshifts are shown with filled circles while sources with photometric redshifts are shown with open circles. }\label{lir}
\end{figure*}

We integrate the best fit template over 8 -- 1000 $\mu$m to derive the total IR luminosity (L$_{IR}$) (Figure \ref{lir}). We find luminous IR galaxies (LIRGs, $10^{11} L_\odot< $ L$_{IR}$ $< 10^{12} L_\odot$) are detected out to $z \sim 1.25$, while ultraluminous IR galaxies (ULIRGs, L$_{IR}$ $> 10^{12} L_\odot$) are detected out to $z$ = 3. 

\subsubsection{X-ray Counterparts}
Using the 2Ms Chandra data \citep{Luo08}, we find 55 of the 70$\,\mu$m sources are within 2 arcseconds of an X-ray source. We use this data to calculate the hardness ratio so as to discriminate against AGN (Section \ref{agndiag}).

\section{Results and Analysis}

We define $q_{IR}$ as 

\begin{equation}
q_{IR} = \log_{10}(\frac{S_{IR}}{S_{radio}}),
\end{equation}
where
$S_{IR}$ is the observed infrared flux density at the specified IR wavelength (e.g. 70$\,\mu$m), and
$S_{radio}$ is the observed flux density at 1.4\,GHz. 

\subsection{AGN identification}\label{agndiag}

We wish to study the FRC of predominantly star-forming galaxies and so we removed galaxies from our sample if they satisfied any of the following four criteria which indicate AGN. 
 
\begin{figure*}
\begin{center}
\begin{tabular}{ccc}
\includegraphics[angle=-90, scale=0.22]{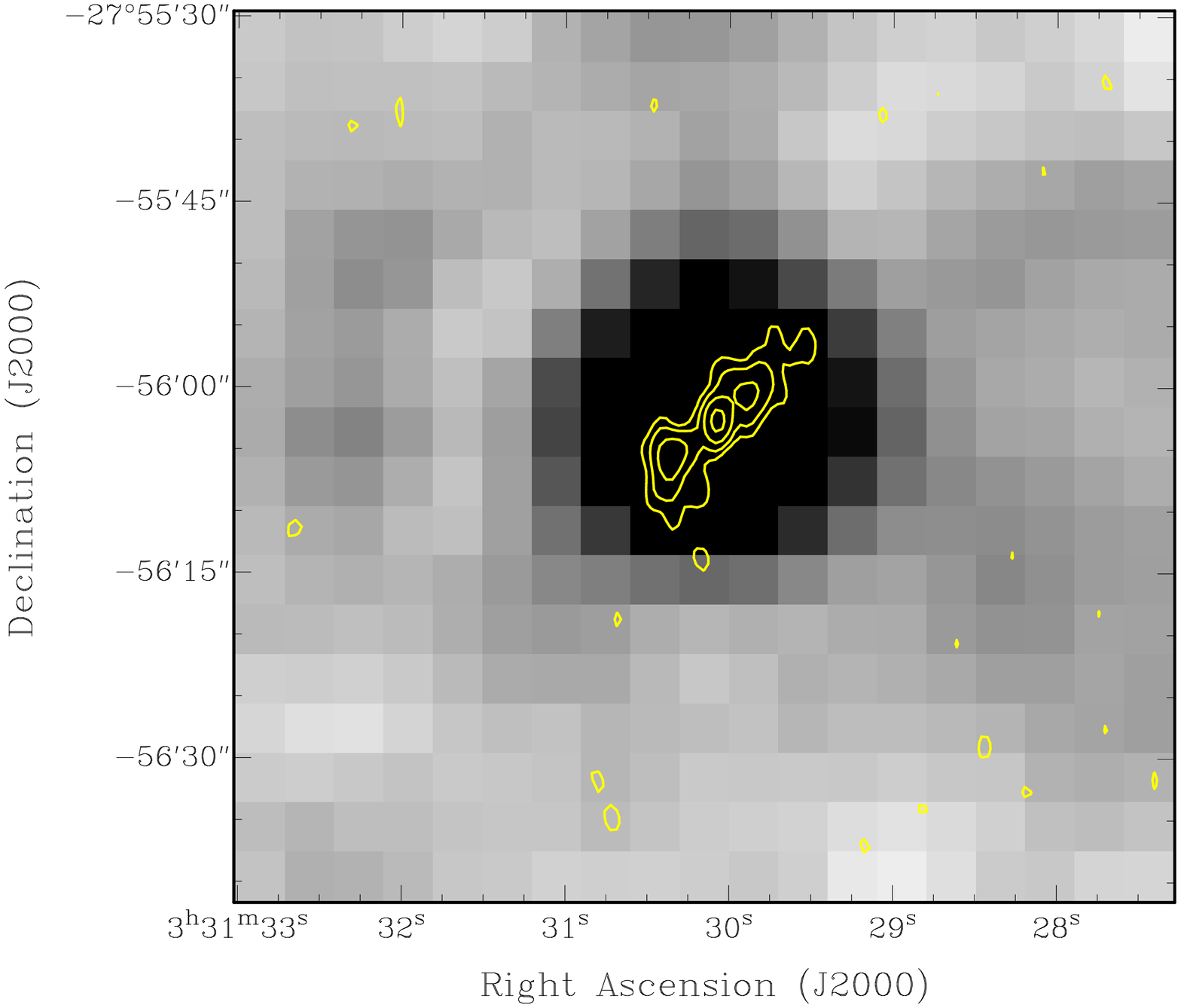} & \includegraphics[angle=-90, scale=0.22]{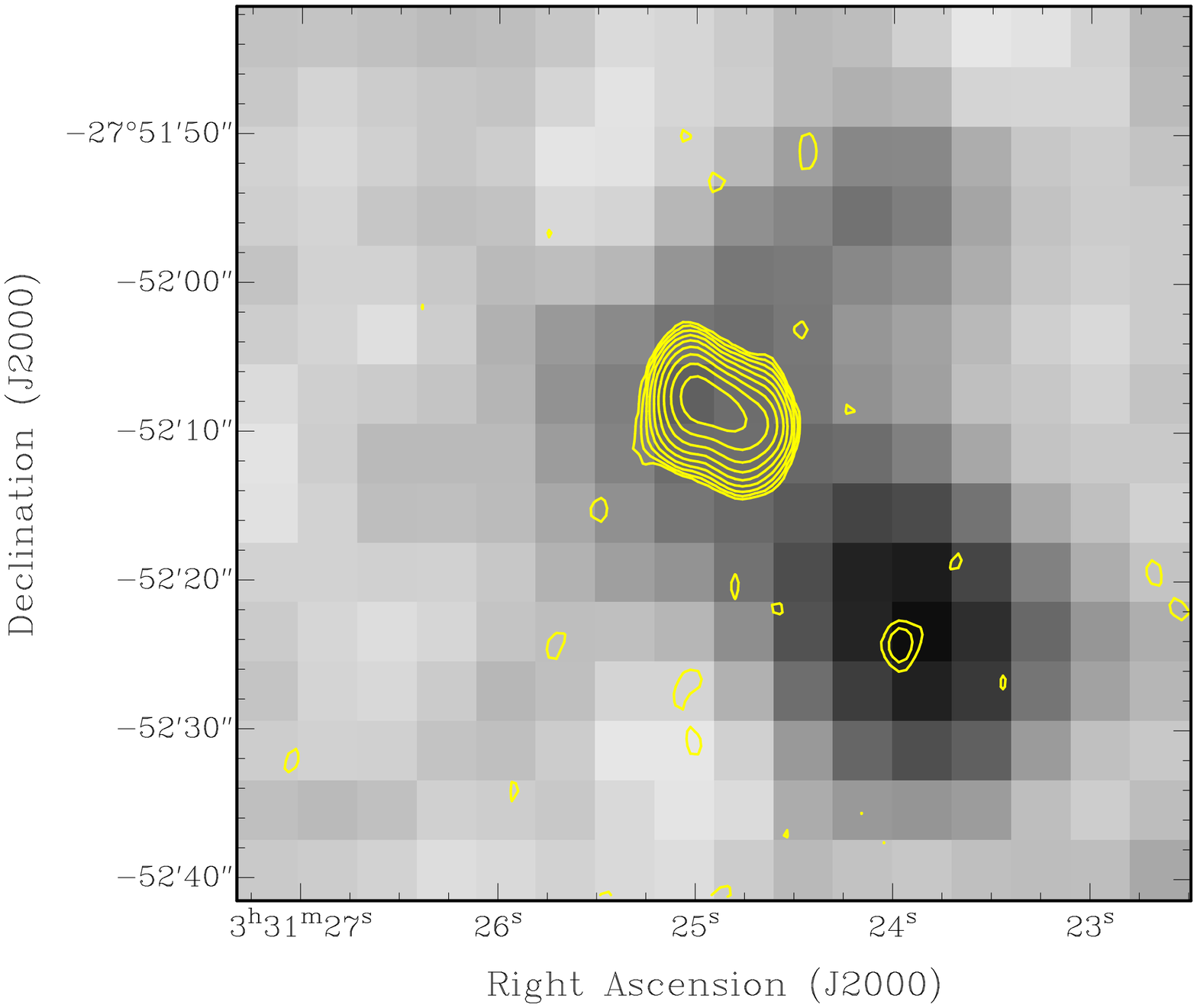} & \includegraphics[angle=-90, scale=0.22]{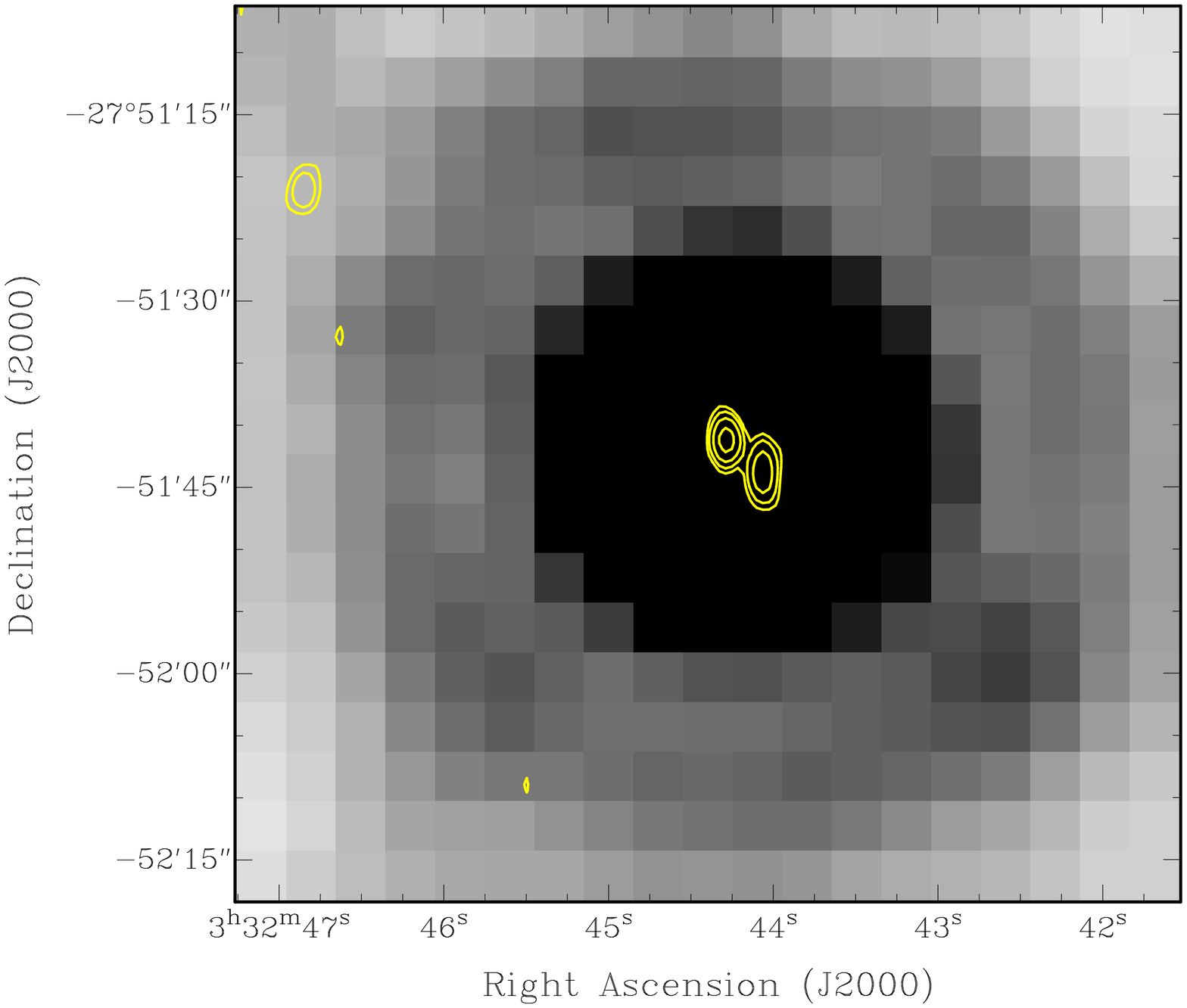}\\
\end{tabular}
\end{center}
\caption{Sources which we classified as AGN based on radio morphology. The greyscale image is the 70$\,\mu$m image and contours are the 1.4\,GHz image starting at 24 $\mu$Jy ($\sim$3 times the rms) and increasing by factors of 2.}\label{radiomorph}
\end{figure*}

\begin{figure*}
\begin{center}
\includegraphics[angle=0, scale=0.8]{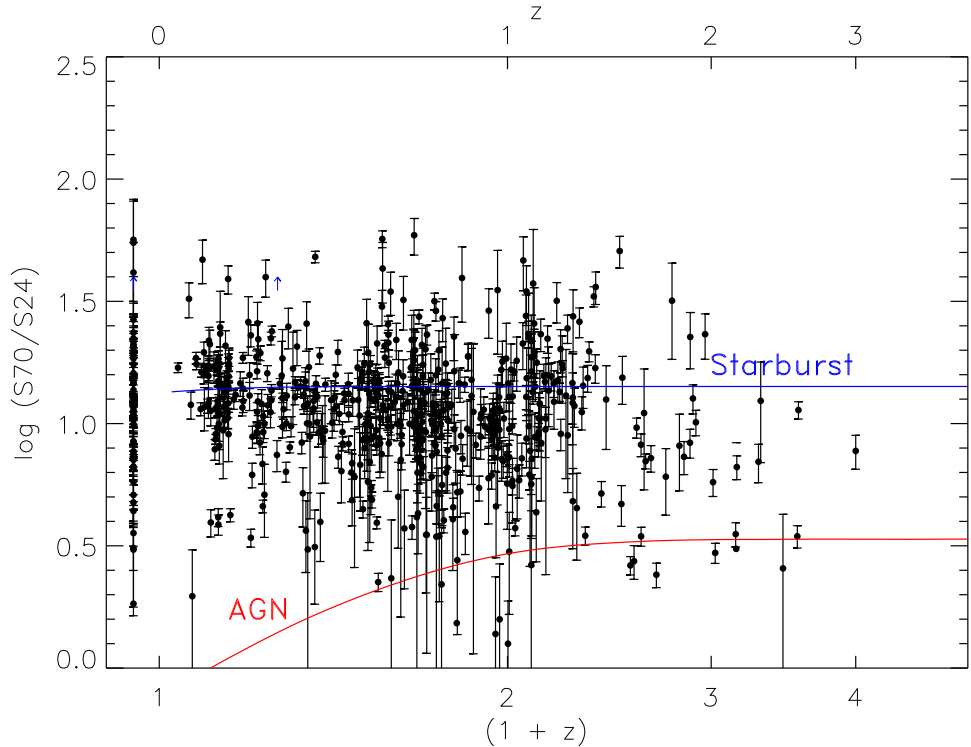}
\end{center}
\caption{The ratio of 70$\,\mu$m flux density over 24$\,\mu$m flux density plotted against redshift. Sources with no redshift information are shown at an artificial redshift of -0.05. The blue line represents a simple modified blackbody SED model for starbursts with a dust temperature of 30K with a mid-infrared slope of $\alpha$ = 2.4, while the red line represents a model for AGNs with a dust temperature of 90K with a mid-infrared slope of 1.1 \citep{Frayer06}. Error bars are derived by combining in quadrature the standard errors in radio and IR fluxes.}\label{s70s24}
\end{figure*}

\begin{enumerate}
\item{The source has a radio morphology that displays the classic double-lobed AGN morphlogy. Figure \ref{radiomorph} shows the three sources that were identified and subsequently removed from our sample. }
\item{The source has log(S70/S24) $<$ 0.5, because AGNs are expected to have low S70/S24 ratios \citep{Frayer06}. 22 sources were identified as AGN in this way. Figure \ref{s70s24} shows the log of the ratio of 70$\,\mu$m flux density over 24$\,\mu$m flux density, plotted against redshift. }
\item{The source has a hardness ratio $>$ 0.2 based on the 2Ms Chandra data \citep{Luo08}, where hardness ratio is defined as:
\begin{equation}
HR = \frac{S_{hard} - S_{soft}}{S_{hard} + S_{soft}},
\end{equation}
where
$HR$ is the hardness ratio,  
$S_{hard}$ is the flux density of the 2 - 8 keV band,
$S_{soft}$ is the flux density of the 0.5 - 2.0 keV band \citep{Rosati02}. 14 sources were identified as AGN in this way.}
\item{The source has a soft X-ray (0.5-2 keV) to R-band flux density ratio of greater than one, as used by \citet{Luo08} to classify AGN. 14 sources were identified as AGN in this way.}
\end{enumerate}

Low $q_{70}$ and $q_{24}$ values may also be used to discriminate against AGN \citep[e.g.][]{Middelberg08}. However, doing so could potentially bias our results against low values of $q$. Applying these criteria removed only a further five AGN from our sample and we found made a negligible difference to our results, so we do not use these criteria here.

In total, 44 ($\sim$7$\%$) sources were classified as AGN and subsequently removed from further analysis, leaving us with a final catalog of 573 70$\,\mu$m sources. Most of the sources were identified as AGN based on only one of the above four diagnostics, highlighting the need for multiple AGN diagnostics. There were only six sources that were identified as AGN by both the X-ray diagnostics. 41/44 of the sources identified as AGN had redshift information. The median redshift of the AGNs is 0.969, slightly higher than the median redshift of the entire sample (0.655). The median L$_{IR}$ of the AGNs is 5.10 $\times$ 10$^{11}$ L$_{\odot}$, also higher than the median L$_{IR}$ of the entire sample (2.07 $\times$ 10$^{11}$ L$_{\odot}$). The AGN tend to be at higher redshifts and higher L$_{IR}$ due to Malmquist bias.

\subsection{$q_{70}$}\label{secq70}

\begin{figure*}
\begin{center}
\includegraphics[angle=0, scale=0.8]{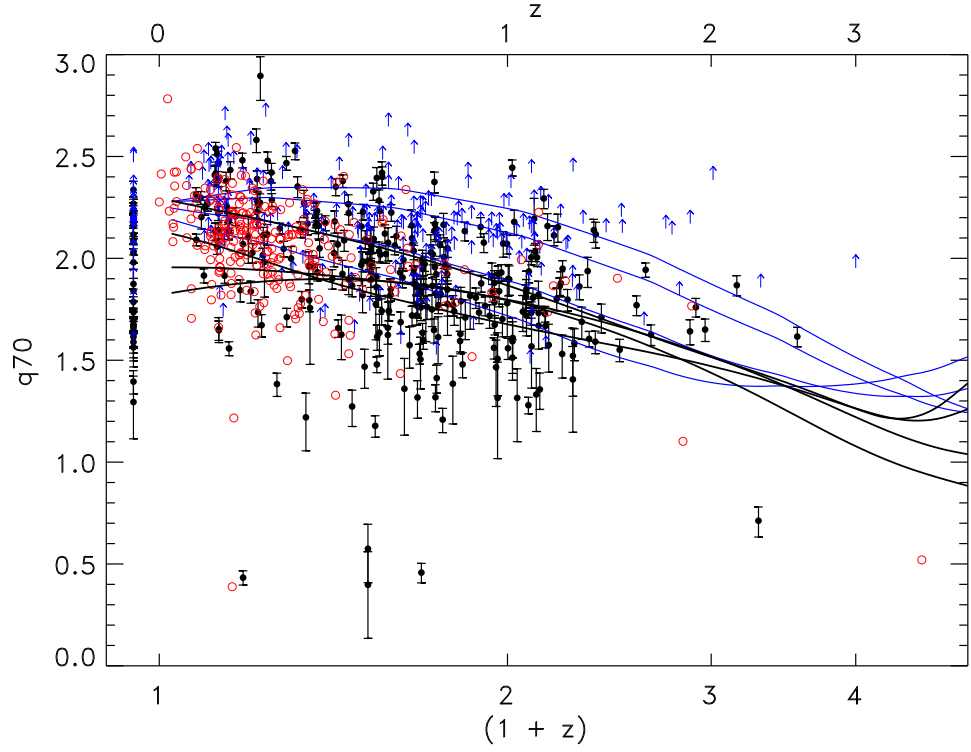}
\end{center}
\caption{The 70$\,\mu$m FIR-radio correlation plotted against redshift. The black points are sources that have both 70$\,\mu$m and 1.4\,GHz detections while the blue arrows show the lower limit of $q_{70}$ for sources with no radio detection. The red open circles are data from xFLS \citep{Appleton04}. The black lines are expected q$_{70}$ tracks derived from SED templates for galaxies with total infrared luminosities of 10$^{9}$ (normal galaxies), 10$^{11}$, 10$^{12}$ and 10$^{13}$L$_{\odot}$ (ULIRGs) going from bottom to top, from \citet{Chary01} and the blue lines are q$_{70}$ tracks derived from SED templates from \citet{Dale02}. Sources with no redshift information are shown at an artificial redshift of -0.05. Error bars are derived by combining in quadrature the standard errors in radio and IR fluxes. }\label{q70}
\end{figure*}

To explore how the FRC, as probed by the observed 70$\,\mu$m band, changes with redshift, we plot $q_{70}$ against redshift in Figure \ref{q70}. Our data are plotted in black and lower limits of $q_{70}$ from the radio non-detections are plotted as blue arrows. The lower limits are calculated by choosing the upper limit to the radio flux density for the radio non-detections to be 3 $\times$ rms. Data from the Spitzer Extragalactic First Look Survey \citep[xFLS,][]{Appleton04} are plotted as red open circles. It is immediately evident that our data contain many more high redshift ($z$ $>$ 1) sources than the xFLS data. \citet{Appleton04} found a $q_{70}$ value of 2.15 $\pm$ 0.16 for their sample. The data points that are at low $q_{70}$ values are likely to be AGN.

We also overplot the expected $q_{70}$ ratio as a function of redshift from IR SED templates of \citet{Dale02} (DH02) and \citet{Chary01} (CE01). The CE01 SED templates are obtained from low-redshift galaxies with appropriate luminosities. We use four CE01 templates which range in L$_{IR}$ from 10$^9$ (normal galaxy) to 10$^{13}$\,L$_{\odot}$ (ULIRGs). The DH02 SED templates are derived from combining theoretical SEDs from dust emission of individual regions within a galaxy. We use four DH02 templates with $\alpha$= 1, 1.5, 2 and 2.5, where $\alpha$ represents the relative contributions of the different dust emission SEDs. The overplotted $q_{70}$ tracks are derived using the IR SED templates, while the radio flux is derived from the infrared flux assuming both thermal and non-thermal components and a constant value of q. 

\begin{figure*}
\begin{center}
\includegraphics[angle=0, scale=0.8]{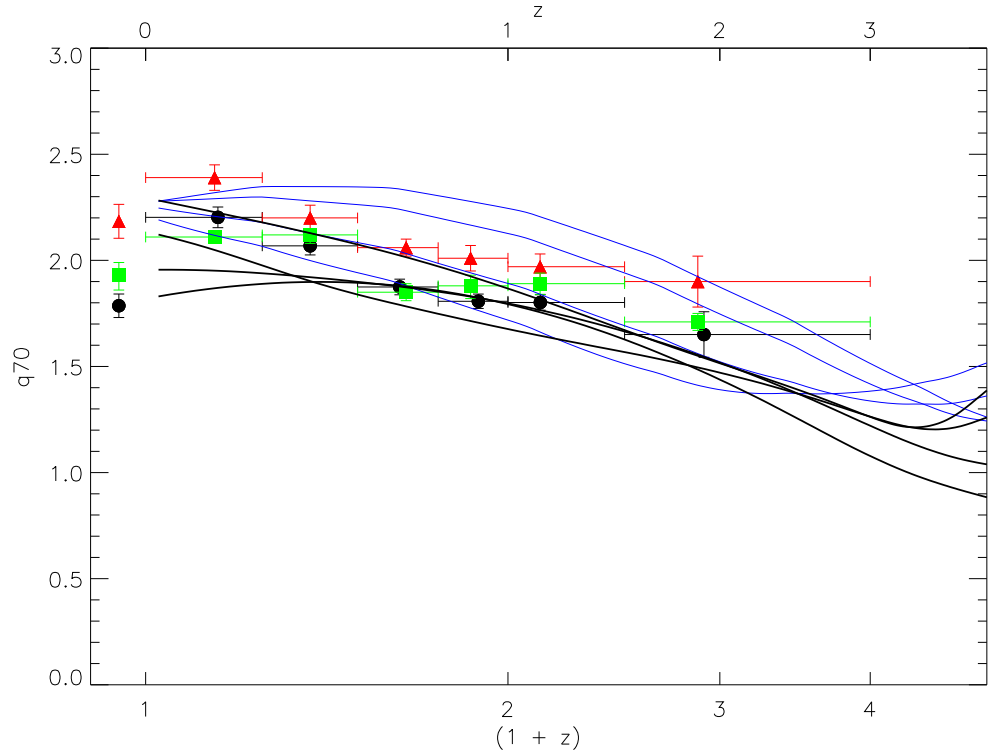}
\end{center}
\caption{Median $q$ values for different redshift bins. The black circles show the median $q_{70}$ values for sources with both 70$\,\mu$m and 1.4\,GHz detections, the red triangles show the median $q_{70}$ value for all sources taking into account the lower limits using survival analysis and the green squares show the median $q_{70}$ value after stacking. The tracks derived from the SED templates are the same as in Figure \ref{q70}. Vertical error bars denote standard errors and horizontal error bars indicate the range of the redshift bin. }\label{medqplot}
\end{figure*}

In order to quantify the evolution of $q_{70}$, we binned all data with both 70$\,\mu$m and radio detections in redshift and determined the median $q_{70}$ in each redshift bin. The results are shown in Table \ref{medq_detect} as well as in the black points of Figure \ref{medqplot}. We calculate a median $q_{70}$ of 2.16 $\pm$ 0.03 out to $z$ = 0.5, which is in agreement with \citet{Appleton04}, as well as other previous studies \citep[e.g.,][]{Seymour09}. Sources with no redshift information have a mean $q_{70}$ ratio similar to $z$ ~$\sim$ 1 sources. 

Figure \ref{medqplot} shows no evidence for evolution in the FIR-radio correlation because the data points are consistent with the overplotted $q_{70}$ tracks. At $z$ $>$ 1.25 we are detecting only ULIRGs (see Figure \ref{lir}), and this is reflected in the higher median $q_{70}$ values, which are better traced by the $q_{70}$ track derived from the ULIRG SED template.

\subsection{Survival Analysis}\label{surv}

Our 70$\,\mu$m selected sample has sources with both 70$\,\mu$m and 1.4\,GHz detections, as well as sources that do not have 1.4\,GHz detections. To study the full IR sample, we have to include radio non-detections. As we know the rms of the 1.4\,GHz data we can estimate an upper limit to the radio flux density, which translates to a lower limit for $q_{70}$. We choose the upper limit to the radio flux density for the radio non-detections to be 3 $\times$ rms. One method of including limits from non-detections is using a branch of statistics called survival analysis. Survival analysis is an extensive field of statistics and was first applied to astronomy by \citet{Feigelson85}. We perform survival analysis on our data using the ASURV (Astronomical SURVival analysis) package developed by \citet{Lavalley92}. 

We took the 50$^{th}$ percentile of the Kaplan-Meier estimator to obtain a median $q_{70}$ value for each redshift bin. The results are shown in Table \ref{medq_all}, as well as in Figure \ref{medqplot}. The two highest redshift bins appear to have higher median $q_{70}$ values, but this is because at $z$ $>$ 1.25 we are detecting only ULIRGs, and hence these two points lie along the most actively star-forming, or ULIRG track. 

Our $q_{70}$ values appear to agree, within the errors, to the $q_{70}$ tracks derived from the empirical SED templates from the local Universe. This is evidence for little, if any evolution in $q_{70}$, which implies that galaxies at high redshifts ($z$ $\sim$ 2.5) share many of the same properties as galaxies in the local Universe. 

\subsection{Stacking}\label{stacking}

\begin{figure*}[bt]
\includegraphics[width=2.2cm]{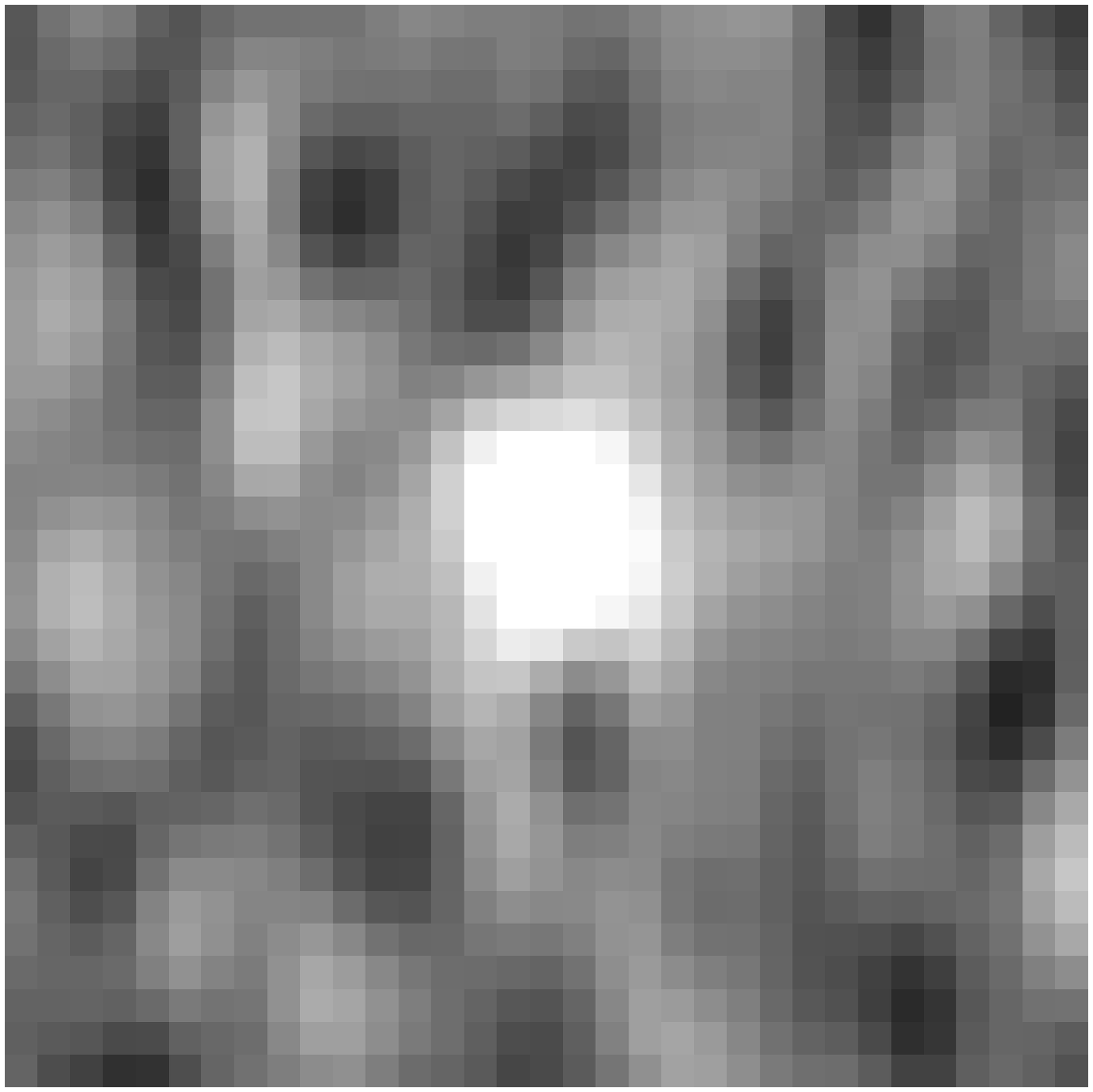}
\includegraphics[width=2.2cm]{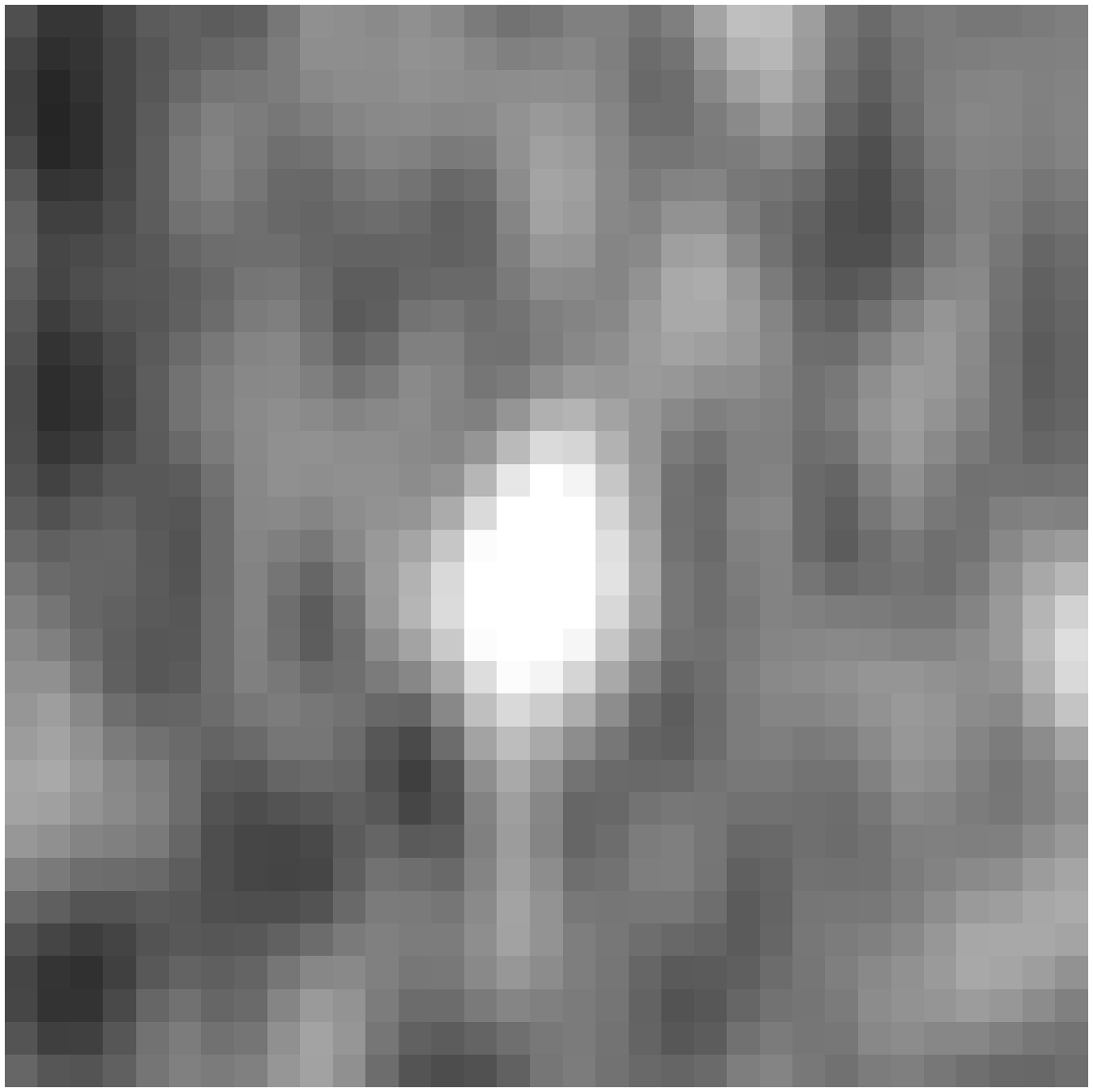}
\includegraphics[width=2.2cm]{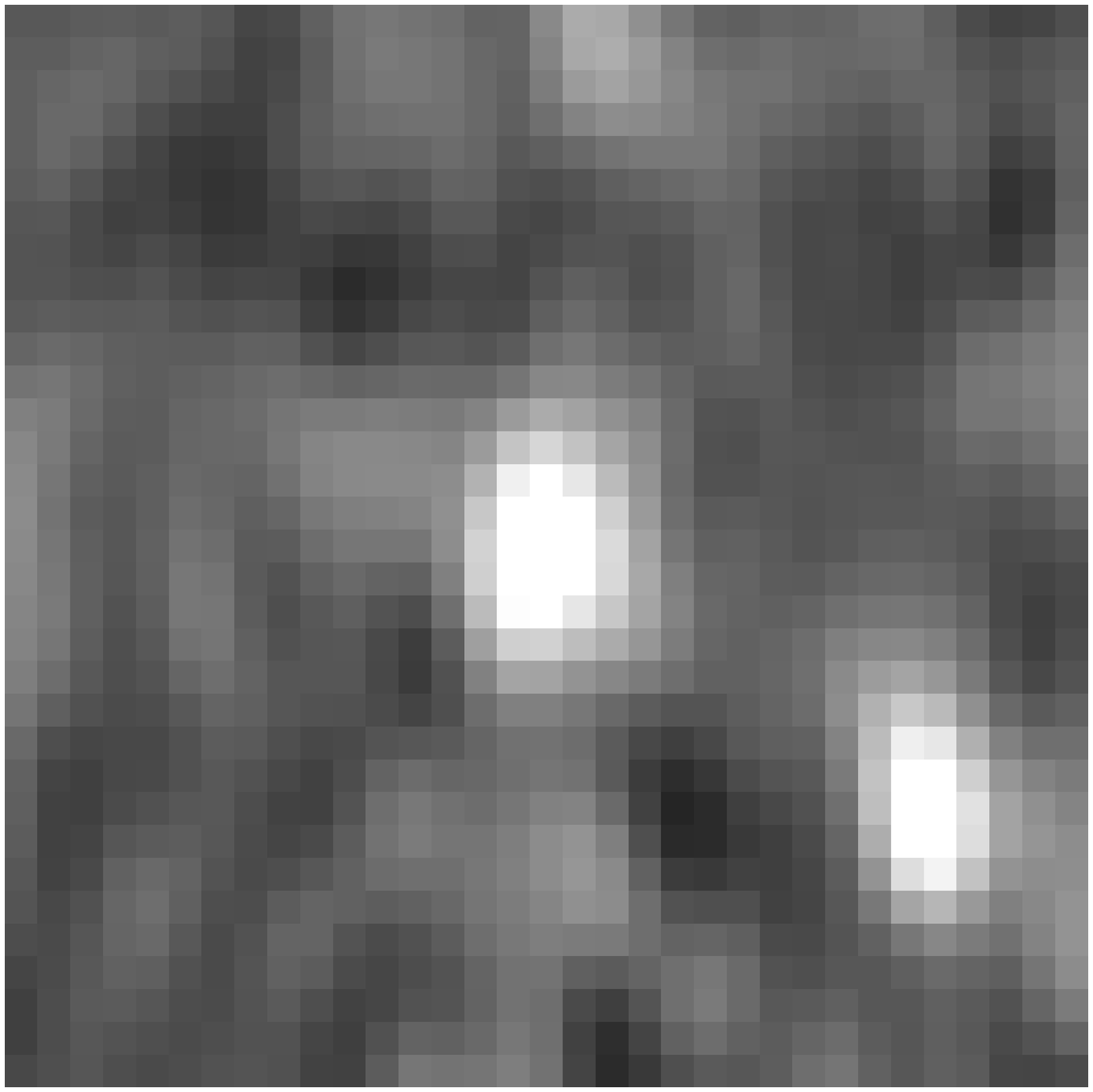}
\includegraphics[width=2.2cm]{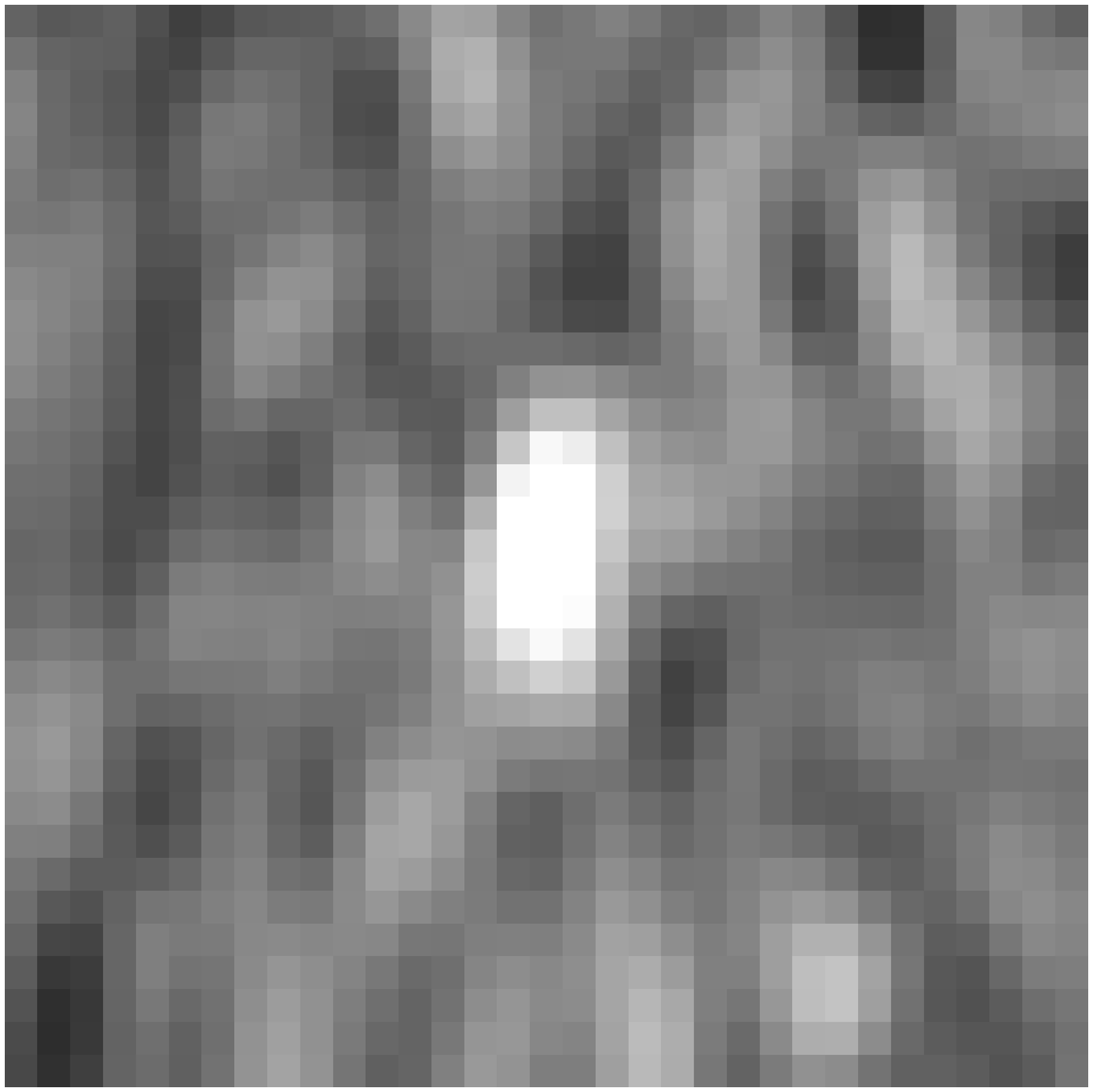}
\includegraphics[width=2.2cm]{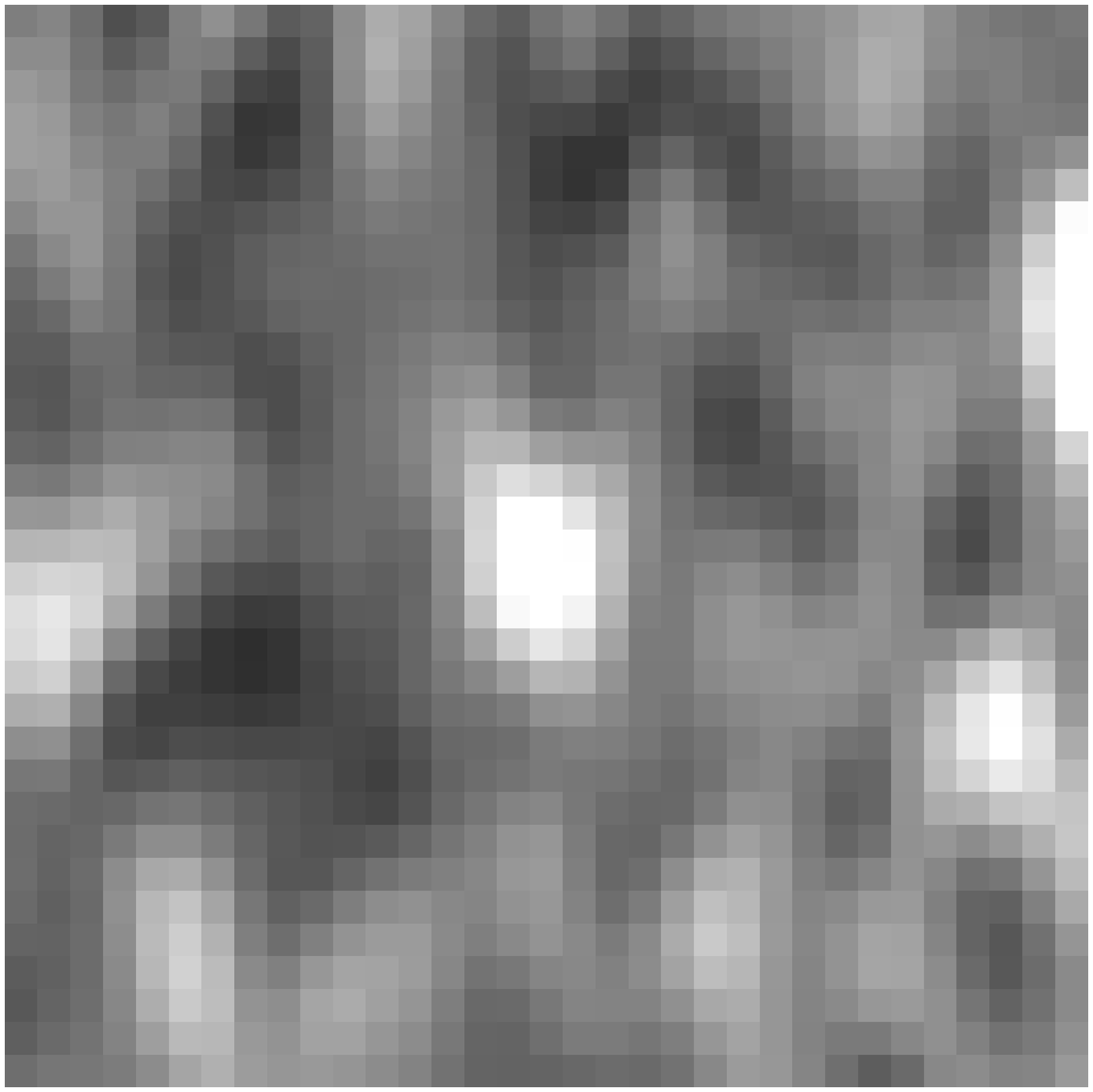}
\includegraphics[width=2.2cm]{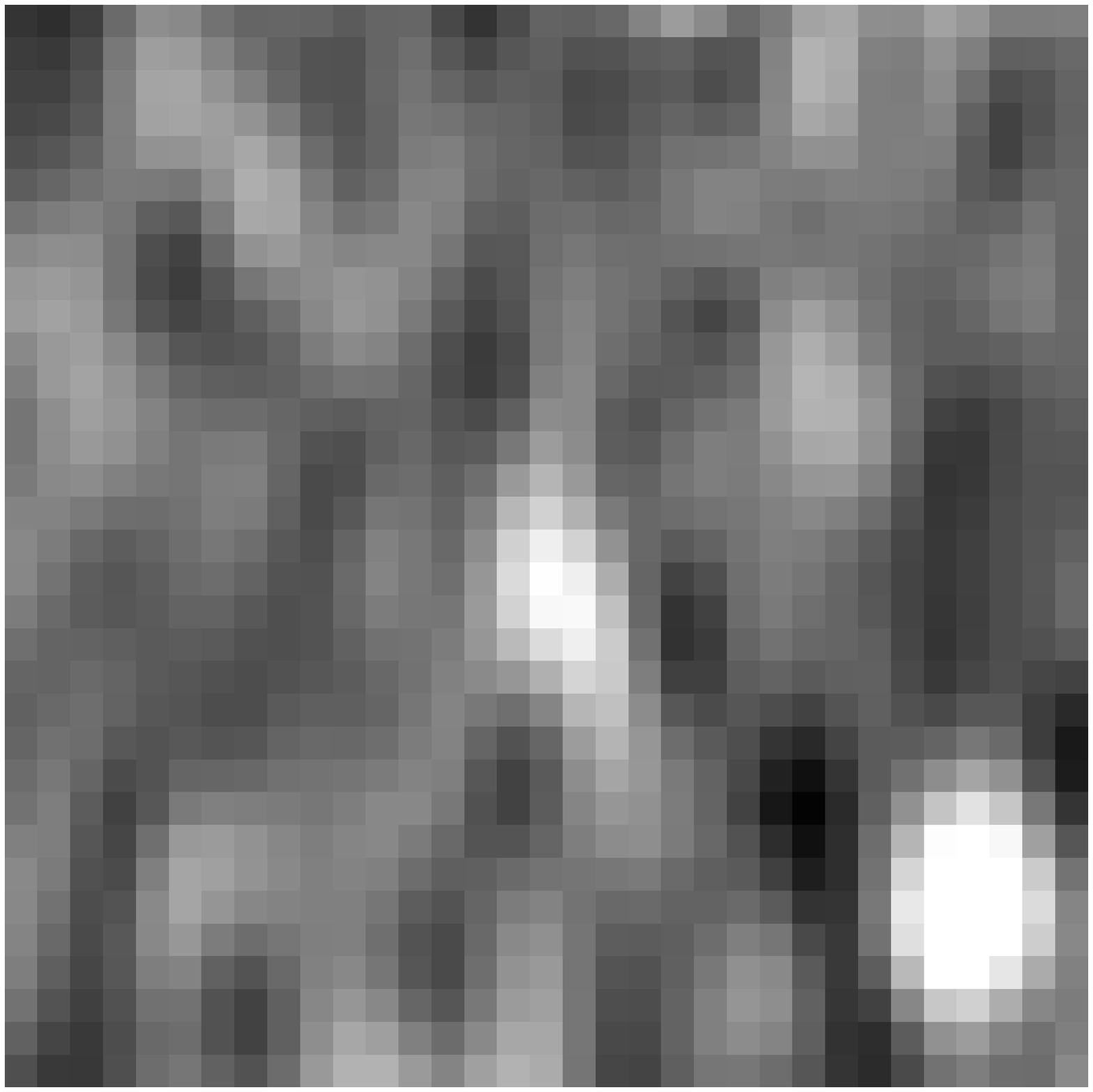}
\includegraphics[width=2.2cm]{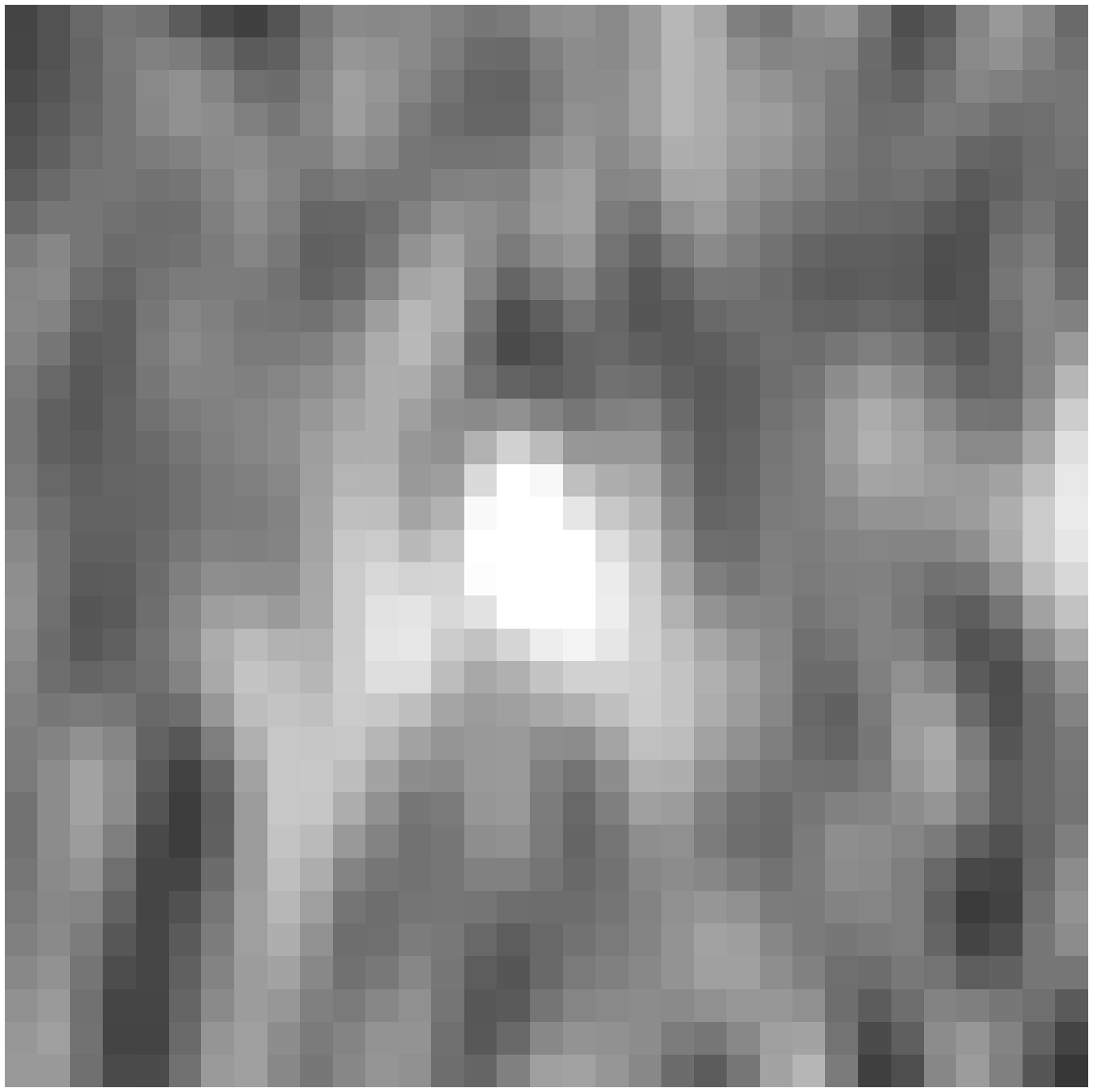}
\vspace*{-2mm}
{\tiny 
\begin{tabbing}
\hspace*{0.25cm} \= \hspace*{2.1cm} \= \hspace*{2.25cm} \= \hspace*{2.25cm}   \= \hspace*{2.25cm}
  \= \hspace*{2.65cm}  \= \hspace*{2.5cm}  \= \hspace*{1.9cm}
 \kill\> 
$0 \le z < 0.25$ \> $0.25 \le z < 0.50$  \>   $0.50 \le z < 0.75$ \> $0.75 \le z < 1.00$ \> $1.00 \le z
 < 1.50$ \>  $z \ge 1.5$ \> no $z$\\
\end{tabbing}
}
\caption{Postage stamps of the radio stacks of the IR sources, for the redshift bins as shown. 
The stacks are  16 $\times$ 16 arcsec in size. There is significant radio flux at the center of the stack in all of the redshift bins. The measured flux densities are reported in Table \ref{stacktable}.}\label{stacks}
\end{figure*}

To extend the results we performed a stacking analysis of the IR sources by stacking the radio sources in bins of redshift. We started by stacking the M08 radio data at the positions of all sources not detected in the radio image, using a noise-weighted mean, for which the noise was determined from the radio image in the vicinity of the IR source. The stacked radio flux density was then obtained from a Gaussian fit to the signal at the center of the radio stack, leaving all Gaussian parameters unconstrained. The resulting stacks are shown in Figure \ref{stacks}, and each has a significant detection.

To derive an average $q_{70}$ for all IR sources in the various redshift bins (green points in Figure \ref{medqplot}), we calculated a mean by adding the flux densities of the detected sources to the stacked flux density. Each was added with a weight =1/n (where n is the total number of measurements, including both stacked and detected images) and the stacked flux was weighted by m/n where m is the number of stacked images. The results are summarized in Table \ref{stacktable}. The higher redshift bins have higher $q$ values when compared to the survival analysis results because stacking analysis imposes lower limits on the radio flux densities. As before, the FIR-radio flux density ratios appear to follow the ratios expected for local SEDs and hence we find little evolution in the FIR-radio correlation as a function of time.

\subsection{$q_{TIR}$}

We computed the ratio of total infrared luminosity over rest frame radio luminosity to confirm our results from Sections \ref{secq70} to \ref{stacking}. Furthermore, this allows us to make direct comparisons with other studies that compute the FRC in this way. 

We computed $q_{TIR}$ for all 521 sources that had redshift information (and were not identified as AGN) using:

\begin{equation}
q_{TIR}=\log(\frac{L_{IR}}{3.75 \times 10^{12}{\rm W}}) - \log(\frac{L_{1.4\,GHz}}{{\rm W Hz}^{-1}}),
\end{equation}
where
$L_{IR}$ is the total infrared luminosity (Section \ref{zlir}), and
$L_{1.4GHz}$ is the rest-frame 1.4\,GHz luminosity.  In order to account for k-correction in the radio, we set the spectral index, $\alpha$, to 0.8\footnote{S$_{\nu} \propto \nu^{-\alpha}$, where S$_{\nu}$ is the flux density at frequency $\nu$.}, following the work of \citet{Ibar10} and \citet{Sargent10b}.

\begin{figure*}
\begin{center}
\includegraphics[angle=0, scale=0.8]{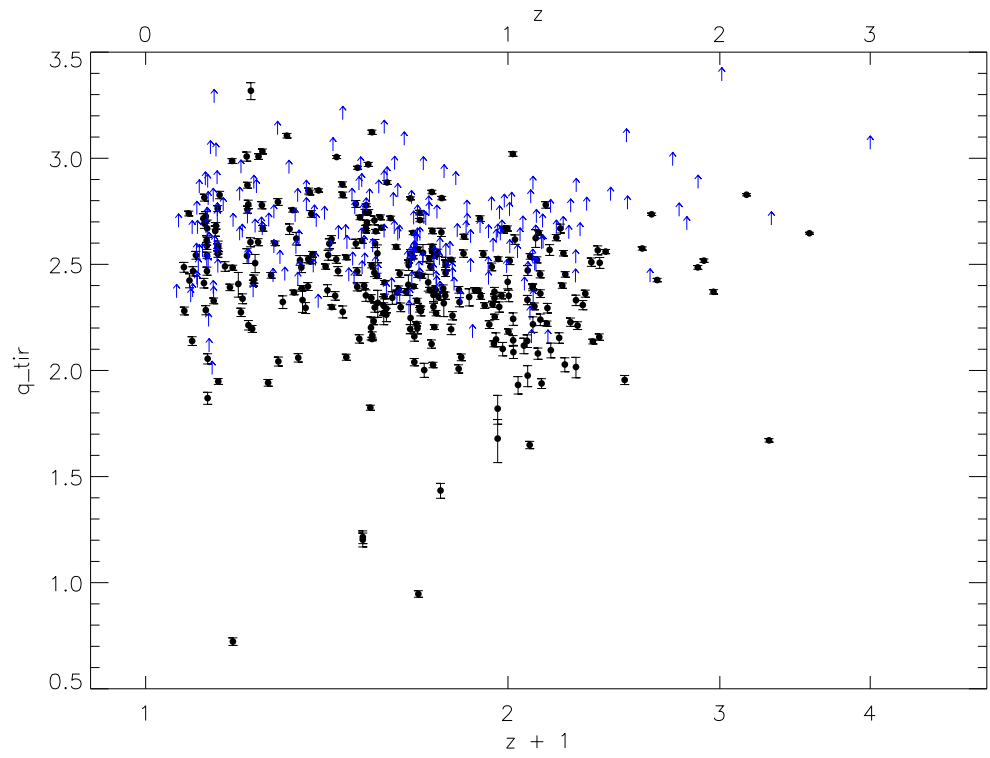}
\end{center}
\caption{The total infrared luminosity FIR-radio correlation plotted against redshift. The same symbol and color scheme is used as for Figure \ref{q70}. }\label{qtirvz}
\end{figure*}

\begin{figure*}
\begin{center}
\includegraphics[angle=0, scale=0.8]{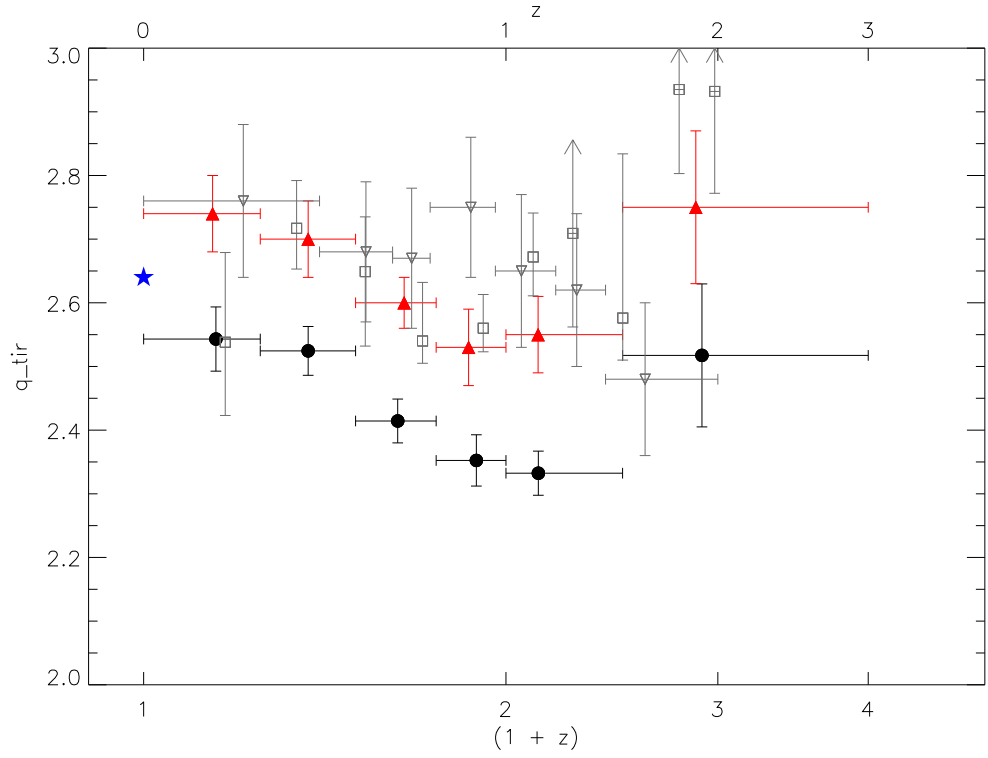}
\end{center}
\caption{Median $q_{TIR}$ values for different redshift bins. The black filled circles show the median $q_{TIR}$ values for sources with both 70$\,\mu$m and 1.4\,GHz detections and the red triangles show the median $q_{TIR}$ value for all sources taking into account the lower limits using survival analysis. The grey upside-down open triangles show the median $q_{TIR}$ values derived by \citet{Bourne10} and the grey open squares show the median $q_{TIR}$ values derived by \citet{Sargent10b} for star-forming galaxies. None of the data shown include the Kellermann correction (Section \ref{kellcorr}). Vertical error bars are standard errors for both our dataset and Bourne's data, but the vertical error bars for Sargent's data are upper and lower 95\% confidence levels. The blue star at (1+$z$) = 1 represents the median $q_{TIR}$= 2.64 $\pm$ 0.02 from \citet{Bell03}.}\label{medqtirvz}
\end{figure*}

Figure \ref{qtirvz} shows the distribution of $q_{TIR}$ as a function of redshift. For sources without a radio detection we set an upper limit to the radio flux of 3 $\times$ rms. Similar to Figure \ref{q70}, sources without a radio detection (ie. lower limits) tend to have higher $q_{TIR}$ when compared to sources with a radio detection. In order to take into account the effect of the radio non-detections,  we bin our data in redshift and perform survival analysis. Figure \ref{medqtirvz} shows the median $q_{TIR}$ for only the detected sources (black circles), and for all sources by using survival analysis (red triangles). We have included results from previous studies in Figure \ref{medqtirvz}. The blue star at $z$ = 0 is at $q_{TIR}$ = 2.64, the mean $q_{TIR}$ found by \citet{Bell03}. The open upside-down triangles are $q_{TIR}$ derived from stacking analyses by \citet{Bourne10} and the open squares are the ``uncorrected'' $q_{TIR}$ for star-forming galaxies, from \citet{Sargent10b} (See Section \ref{kellcorr} for a discussion on the Kellermann Correction).

 Our $q_{TIR}$ values are all within $\sim$0.22 of each other. Our values agree, within the errors, with the work of \citet{Sargent10b}, and also agree with the work of \citet{Bourne10}, with the exception of the redshift bins 0.75 $<$ z $<$1 and z $>$ 1.5, where our values differ by $<$2$\sigma$ compared to the values given by \citet{Bourne10} for similar redshift ranges. \citet{Sargent10b} use 3 $\times$ rms to determine the upper limit of the radio flux density for their dataset. Unlike the work by \citet{Bourne10} and \citet{Sargent10b}, our $q_{TIR}$ values do not consistently increase or decrease over the redshift range studied, although we do see marginally significant evidence for a decreasing $q_{TIR}$ over the redshift range 0 to 1.5, followed by an increase from 1.5 to 3. However, Fig 14 shows that the data points that use survival analysis (which are more reliable than those that do not) are statistically consistent with a horizontal line at about $q_{TIR}$ = 2.62, and so we conclude that these apparent variations of $q_{TIR}$ with redshift  may not be statistically significant.


\citet{Bourne10} find their $q_{TIR}$ values are systematically higher than the median $q_{TIR}$ found by \citet{Bell03} of 2.64 $\pm$ 0.02, except at $z$ $>$ 1 where it appears to decline. They suggest the apparent decline may be attributed to the assumptions they make about spectral indices.  \citet{Sargent10b} find that their $q_{TIR}$ values are constant with redshift when they apply a correction (see Section \ref{kellcorr}), otherwise $\sim$0.3 dex of evolution is found. While \citet{Bourne10} also uses FIDEL data in the ECDFS, their sample is IRAC selected whereas our sample is 70$\,\mu$m selected. \citet{Sargent10b} use a 24$\,\mu$m selected sample to study the $q_{TIR}$ in the COSMOS field. 

There are a number of factors that may contribute to uncertainties in $q_{TIR}$. The predominant source of uncertainty is in the estimation of $L_{IR}$ as we are fitting the 24\,$\mu$m and 70\,$\mu$m fluxes to SED templates. The uncertainty from the choice of model SEDs can add 0.2 to 0.3 dex \citep{Magnelli09, LeFloch05}. Uncertainties also arise from the rest-frame radio luminosity where we are assuming $\alpha$ = 0.8. 

\citet{Jarvis10} suggest that the slight upturn in $q_{IR}$ seen at low redshifts ($z$ $<$ 0.5) in high redshift studies of the FRC \citep[e.g.][]{Bourne10} may be attributed to the resolving  of extended structure, which means radio flux would be missed and hence increase the FRC. Our radio data has a 2.8$''$ $\times$ 1.6$''$ beam, so  if the Jarvis hypothesis is correct, we would expect to see this effect at low redshifts. 

Although we see a slight upturn in $q_{TIR}$ from $z$ = 1 to $z$ = 0, this increase has a roughly constant gradient, implying that resolution effects are important as high as $z$ = 1. However, at $z$ = 1 a typical star-forming galaxy of diameter 10 kpc is unresolved by our beam. Thus, the increasing $q_{TIR}$ towards low redshifts is probably not due to this effect.

\subsubsection{The Kellermann Correction}\label{kellcorr}

\citet{Kellermann64} showed that flux ratios (such as q) or spectral indices of a flux-limited sample are biased by a factor which depends sensitively on the scatter in the flux ratio and on the source intensity distribution, and in particular on the power law index $\beta$ of the differential source counts (i.e., dN/dS $\propto$ S$^{-\beta}$ ). Other formulations of the same effect are given by \citet{Condon84}, \citet{Francis93} and \citet{Lauer07}, and were first noted as being relevant to the evolution of $q$ by \citet{Sargent10a}.

\citet{Sargent10b} assume a Euclidean ($\beta$=2.5) source intensity distribution at $z$ $<$ 1.4 and a sub-Euclidean ($\beta$ $\sim$1.5) at 1.4 $<$ $z$ $<$ 2 to derive a correction of 0.22 in their value of $q$ at $z$ $>$ 1.4. However we note that (a) the expression assumes a common value of $\beta$ for both the radio and infrared source counts, and (b) the value of $\beta$ is very uncertain at low flux densities. For example, the median radio flux density for our detected sources at $z$ $<$ 1.4 is $\sim$ 72 $\mu$Jy, at which the radio source count power law index lies in the range $\sim$ 1.5 - 2.5 \citep[e.g.,][]{Huynh05}, leading to a correction factor between 0.22 (the adopted value of Sargent, which gives no evolution of $q$ with z) and 0.51 (which implies that $q$ decreases with $z$ with marginal significance).

Because of this uncertainty in the value of the Kellermann correction for faint sources such as those studied here and for consistency with other authors \citep[e.g.,][]{Appleton04, Seymour09, Bourne10}, we do not apply it to our data but note that it is responsible for a further uncertainty in the slope of $q$ as a function of redshift. The range of possible Kellermann corrections (which depend strongly on the assumed value of $\beta$) to the variation of $q$ over our redshift range is roughly centered on zero, with a  total spread of about $\pm$ 0.3.

\section{Summary and Conclusions}

We have studied the FRC out to $z$ $>$ 2  of ULIRGs in ECDFS.  Our results for $q _{70}$ showed that they could be broadly described by $q_{70}$ tracks derived from the SED templates of \citet{Chary01} and \citet{Dale02}. To quantify the evolution of $q_{70}$ we binned our data in redshift and determined the median $q_{70}$ using both survival analysis and a stacking analysis. Both survival analysis and the stacking analysis gave similar results. We see no clear evidence for evolution in the FRC at 70$\,\mu$m. 

We also calculate the FRC using $L_{IR}$ and $L_{1.4\,GHz}$ and find that evolution in $q_{TIR}$ is constrained within 0.22. Our calculated $q_{TIR}$ appears slightly lower than previous studies but this may merely be due to uncertainties involved in calculating $q_{TIR}$. We also acknowledge the importance of the Kellermann correction but due to uncertainties in the value we do not apply it to our data.

A lack of evolution in the FRC is surprising because it implies that the myriad of effects on which the FRC rely must either also not evolve, or must evolve in such a way so as to preserve the FRC. Effects such as inverse Compton cooling of the electrons \citep{Murphy09}, evolution of the magnetic field strength, and evolution in the SEDs due to dust and metallicity changes should all affect the FRC.

We therefore conclude that either all these factors are insignificant at $z$ $\sim$ 2 or there is a complex interplay between these factors conspiring in the preservation of the FRC at high redshifts. 

Early science results from \emph{Herschel} have already hinted that the FRC shows no evidence for evolution to $z$ = 2 \citep{Jarvis10, Ivison10b}. More results on the FRC from Herschel are expected over the next few years as observations are completed and the data analyzed.  \emph{Herschel} will measure the far-infrared properties of normal galaxies to $z$ $\sim$ 1 and ULIRGs out to $z$ $\sim$ 4. Consequently,  we will be able to study the FRC out to higher redshifts and hence gain a better understanding of the evolution of star-forming galaxies, especially in the high redshift Universe.

\acknowledgments
We wish to thank our anonymous referee whose insightful comments helped improve this paper. We wish to thank Neal Miller for providing the radio data. MYM was supported by the IPAC Visiting Graduate Research Fellowship Program for this work.

\clearpage

\begin{table*}
\begin{center}
\caption{Table of redshift catalogs. Column 1 names the catalog while Column 2 gives the matching radius which was used to match the redshift data to the 70$\,\mu$m data. Columns 3 and 4 give the catalog size and whether the redshift data is photometric or spectroscopic and Columns 5 and 6 give the total number of sources that are matched to the 70$\,\mu$m data as well as the number that are in the final catalog. Column 7 gives the reference for the catalog. There are 562 sources with redshifts, of which 206 are spectroscopic.}\label{redshifts}
{\footnotesize
\begin{tabular}{ccccccc}
\\
\hline
Catalog & Radius (arcsec) & Spec/Phot & Size & $N_{matched}$ & $N_{final}$ & Reference\\
\hline
COMBO-17 & 2 & phot & 62337 & 555 & 122 & \citet{Wolf04}\\
MUSYC & 1.5 & phot & 59693 & 410 & 234 & \citet{Cardamone10}\\
MUSYC & 1.5 & spec & 2551 & 114 & 26 & \citet{Cardamone10}\\
GOODS & 2 & spec & 7332 (5528) & 160 & 148 & \citet{Balestra10}\\
ATLAS & 5 & spec & 254 & 32 & 32 & \citet{Mao09}\\
\hline
\end{tabular}}
\end{center}
\end{table*}

\begin{table*}
\begin{center}
\caption{Median $q_{70}$ for different redshift bins for only sources with both 70um and radio detections, and are not classed as AGN. There are a total of 323 sources.}\label{medq_detect}
\begin{tabular}{cccc}
\\
\hline
$z$  & median $z$ & $N_{sources}$  & median $q_{70}$\\
\hline

$0 \le z < 0.25$ &     0.148 &          53 &      2.20 $\pm$     0.05 \\
$0.25 \le z < 0.50$  &     0.370 &          46 &      2.07 $\pm$     0.04 \\
$0.50 \le z < 0.75$  &     0.626 &          93 &      1.87 $\pm$     0.04 \\
$0.75 \le z < 1.00$  &     0.890 &          43 &      1.81 $\pm$     0.03 \\
$1.00 \le z < 1.50$  &      1.13 &          52 &      1.80 $\pm$     0.04 \\
$z \ge 1.50$ &      1.91 &           10 &      1.65 $\pm$     0.11 \\
no redshift &   \nodata &          26 &      1.79 $\pm$     0.06 \\

\hline
\end{tabular}
\end{center}
\end{table*}

\begin{table*}
\begin{center}
\caption{Median $q_{70}$ for different redshift bins, calculated using survival analysis, for all 70um sources that are not classed as AGN. Survival analysis was performed for radio flux density limits of 3 $\times$ rms. The range of values given for the $q_{70}$ values derived from model SED templates of DH02 are for $\alpha$ = 2.5 (first value) and $\alpha$ = 1.  The range of values given for the model $q_{70}$ values derived from SED templates of CE01 are for normal galaxies and ULIRGs. There are a total of 573 sources.}\label{medq_all}
\begin{tabular}{cccccc}
\\
\hline
$z$  & median $z$ & $N_{sources}$  & median $q_{70}$ (3$\sigma$)& $q_{70}$ (DH02) & $q_{70}$ (CE01)\\
\hline

$0 \le z < 0.25$ &     0.141 &         107 &     2.39 $\pm$    0.06 &       2.11 - 2.32  & 1.87 - 2.22\\      
$0.25 \le z < 0.50$ &     0.370 &          80 &     2.20 $\pm$    0.06 &       2.00 - 2.35 & 1.90 - 2.13\\   
$0.50 \le z < 0.75$ &     0.646 &         154 &     2.06 $\pm$    0.04 &       1.88 - 2.33 & 1.88 - 2.01\\   
$0.75 \le z < 1.00$ &     0.862 &          75 &     2.01 $\pm$    0.06 &        1.77 - 2.28 & 1.83 - 1.92 \\   
$1.00 \le z < 1.50$ &      1.13 &          85 &     1.97 $\pm$    0.06 &        1.66 - 2.20 & 1.75 - 1.81\\   
$z \ge 1.50$ &            1.88 &           20 &     1.90 $\pm$     0.12 &       1.41 - 1.92 & 1.48 - 1.55\\  
no redshift &           \nodata &          52 &     2.18 $\pm$    0.08 &         & \\   

\hline
\end{tabular}
\end{center}
\end{table*}

\begin{center}
\begin{table*}
\caption{Summary of radio stacking results of IR sources. $N_{det}$ is the number of IR sources in each bin which are detected in the radio image. $N_{stack}$ is the number of IR sources not detected in the radio image and therefore the number stacked in this analysis, resulting in the stacked $S_{1.4\,GHz}$ listed in column 5. Average $q_{70}$ is the FIR-radio flux density ratio for all IR sources in the redshift bin, obtained by combining the detections with the stacked results. }\label{stacktable}
\hfill{}
\begin{tabular}{ccrrcrrr}
\\
\hline 
$z$ & median $z$ & $N_{det}$ & $N_{stack}$ & stacked $S_{1.4\,GHz}$ ($\mu$Jy) & average  $q_{70}$  \\  \hline
$0  \le z < 0.25$ & 0.141 & 53 & 54 & 27.7 $\pm$ 2.8 & $2.11^{+0.02}_{-0.03}$ \\
$0.25  \le z < 0.50$ & 0.370 & 46 & 34 & 25.4 $\pm$ 1.9 & $2.12^{+0.03}_{-0.04}$ \\
$0.50  \le z < 0.75$ & 0.646 & 93 & 61 & 19.7 $\pm$ 1.5 & $1.85^{+0.04}_{-0.04}$ \\
$0.75  \le z < 1.00$ & 0.862 & 43 & 32 & 18.9 $\pm$ 1.6 & $1.88^{+0.05}_{-0.06}$ \\
$1.00  \le z < 1.50$ & 1.13 & 52 & 33 & 13.6 $\pm$ 0.9 & $1.89^{+0.05}_{-0.05}$ \\
$z  \ge 1.50$             & 1.88 &  10 & 10 & 19.5 $\pm$ 2.5 & $1.71^{+0.04}_{-0.04}$ \\
no redshift & \nodata & 26 & 26 & 46.8 $\pm$ 9.4 & $1.93^{+0.06}_{-0.07}$ \\ \hline
\end{tabular} 
\hfill{}

\end{table*}
\end{center}

\begin{table*}
\begin{center}
\caption{Median $q_{TIR}$ for different redshift bins for only sources with both 70$\,\mu$m and radio detections, and are not classed as AGN. There are a total of 297 sources.}\label{medqtir_detect}
\begin{tabular}{cccc}
\\
\hline
$z$  & median $z$ & $N_{sources}$  & median $q_{TIR}$\\
\hline
$0 \le z < 0.25$ &     0.148 &          53 &      2.54 $\pm$    0.05\\       
$0.25 \le z < 0.50$  &     0.370 &          46 &   2.52 $\pm$    0.04 \\      
$0.50 \le z < 0.75$  &     0.626 &          93 &      2.41 $\pm$    0.03 \\   
$0.75 \le z < 1.00$  &     0.890 &          43 &    2.35 $\pm$    0.04 \\     
$1.00 \le z < 1.50$  &      1.13 &          52 &   2.33 $\pm$    0.03 \\      
$z \ge 1.50$ &      1.91 &           10 &         2.52 $\pm$     0.11 \\       

\hline
\end{tabular}
\end{center}
\end{table*}

\begin{table*}
\begin{center}
\caption{Median $q_{TIR}$ for different redshift bins, calculated using survival analysis, for all 70$\,\mu$m sources that are not classed as AGN. Survival analysis was performed using $L_{1.4\,GHz}$ calculated from radio flux density limits of 3 $\times$ rms. There are a total of 521 sources.}\label{medqtir_all}
\begin{tabular}{cccc}
\\
\hline
$z$  & median $z$ & $N_{sources}$  & median $q_{TIR}$ (3$\sigma$)\\
\hline

$0 \le z < 0.25$ &     0.141 &         107 &          2.74 $\pm$    0.06\\
$0.25 \le z < 0.50$ &     0.370 &          80 &         2.70 $\pm$    0.06\\
$0.50 \le z < 0.75$ &     0.646 &         154 &         2.60 $\pm$    0.04\\
$0.75 \le z < 1.00$ &     0.862 &          75 &       2.53 $\pm$    0.06\\
$1.00 \le z < 1.50$ &      1.13 &          85 &      2.55 $\pm$    0.06\\
$z \ge 1.50$ &      1.88 &           20 &           2.75 $\pm$     0.12\\

\hline
\end{tabular}
\end{center}
\end{table*}

\clearpage

%

\newpage

\end{document}